\documentclass{emulateapj}
\usepackage[dvips]{epsfig}
\usepackage{natbib}
\slugcomment{Accepted August 14, 2006}
\newcommand{\cak}{\ion{Ca}{2} K}
\shorttitle{Bright Metal-Poor Stars from the HES I.}
\shortauthors{Frebel et al.}
\begin{document}
\title{Bright Metal-Poor Stars from the Hamburg/ESO Survey. \\
 I. Selection and Follow-up Observations from 329 Fields}
\author{
Anna Frebel\altaffilmark{1},
Norbert Christlieb\altaffilmark{2},
John E. Norris\altaffilmark{1},
Timothy C. Beers\altaffilmark{3},
Michael S. Bessell\altaffilmark{1},\\
Jaehon Rhee\altaffilmark{4},
Cora Fechner\altaffilmark{2},
Brian Marsteller\altaffilmark{3},
Silvia Rossi\altaffilmark{5},
Christopher Thom\altaffilmark{6},\\
Lutz Wisotzki\altaffilmark{7}, and 
Dieter Reimers\altaffilmark{2}
}
\altaffiltext{1}{Research School of Astronomy \& Astrophysics, The Australian
   National University, Cotter Road, Weston, ACT 2611, Australia;
   anna@mso.anu.edu.au, jen@mso.anu.edu.au, bessell@mso.anu.edu.au}

\altaffiltext{2}{Hamburger Sternwarte, Universit\"at Hamburg, Gojenbergsweg
112, D-21029 Hamburg, Germany; nchristlieb@hs.uni-hamburg.de,
cfechner@hs.uni-hamburg.de, dreimers@hs.uni-hamburg.de}  

\altaffiltext{3}{Department of Physics \& Astronomy, CSCE: Center for Study of
Cosmic Evolution, and JINA: Joint Institute for Nuclear Astrophysics, Michigan
State University, E. Lansing, MI 48824-1116, USA; beers@pa.msu.edu,
marsteller@pa.msu.edu}

\altaffiltext{4}{Center for Space Astrophysics, Yonsei University, Seoul
120-749, Korea, and Space Astrophysics Laboratory, California Institute of
Technology, MC 405-47, Pasadena, CA 91125, USA; rhee@caltech.edu}

\altaffiltext{5}{Departamento de Astronomia Instituto de Astronomia,
     Geof{\'\i}sica e Ci{\^e}ncias Atmosf{\'e}ricas, Universidade de S{\~a}o
     Paulo, 05508-900 S{\~a}o Paulo SP, Brazil; rossi@astro.iag.usp.br}

\altaffiltext{6}{Centre for Astrophysics and Supercomputing, Swinburne
University of Technology, Mail $\#31$ PO Box 218, Hawthorn, Victoria 3122,
Australia; cthom@astro.swin.edu.au}

\altaffiltext{7}{Astrophysikalisches Institut Potsdam, An der Sternwarte 16,
D-14482 Potsdam, Germany; lutz@aip.de}

\begin{abstract}
We present a sample of 1777 bright ($9<B<14$) metal-poor candidates selected
from the Hamburg/ESO Survey (HES). Despite saturation effects present in the
red portion of the HES objective-prism spectra, the data were recoverable and
quantitative selection criteria could be applied to select the
sample. Analyses of medium-resolution ($\sim2$\,{\AA}) follow-up spectroscopy
of the entire sample, obtained with several $2$ to $4$\,m class telescopes,
yielded 145 new metal-poor stars with metallicity $\mbox{[Fe/H]}<-2.0$, of
which 79 have $\mbox{[Fe/H]}<-2.5$, and 17 have $\mbox{[Fe/H]}<-3.0$. We also
obtained C/Fe estimates for all these stars. From this, we find a frequency of
C-enhanced ($\mbox{[C/Fe]}>1.0$) metal-poor ($\mbox{[Fe/H]}<-2.0$) giants of
$9\%\pm2\%$, which is lower than previously reported. However, the frequency
raises to similar ($>20\%$) and higher values with increasing distance from
the Galactic plane. Although the numbers of stars at low metallicity are
falling rapidly at the lowest metallicities, there is evidence that the
fraction of carbon-enhanced metal-poor stars is increasing rapidly as a
function of declining metallicity. For $\sim60$ objects, high-resolution data
have already been obtained; one of these, HE~1327$-$2326, is the new record
holder for the most iron-deficient star known.
\end{abstract}


\section{Introduction}
The first systematic searches for metal-poor stars in the Galactic halo, by
means of objective-prism surveys, began with \citet{Bond1970, Bond1980,
Bond1981} and \citet{Bidelman_MacConnell}.  More recent surveys are the HK
survey \citep{BPSI,BPSII,beers_stromlo_symp} and the Hamburg/ESO survey (HES;
\citealt{hespaperIII, Christlieb:2003}). Over time, deeper surveys have became
possible to reach further into the Galactic halo. The limiting magnitude of
the HES is $B\sim 17.5$\,mag as opposed to the earlier HK survey
($B\sim15.5$\,mag).

Particularly the HK survey and HES have both shown that a variety of
astrophysically interesting and unusual objects can be found in a large sample
of metal-poor stars. Some of these exhibit large overabundances of heavy
neutron-capture elements that are synthesized in the s-process (e.g.,
\citealt{2001aokisprocess, 2001vaneck}), or in the r-process (e.g.,
CS~22892$-$052, \citealt{Snedenetal:1996}, or CS~31082-001
\citealt{Cayreletal:2001}). In the case of strong r-process enhancement,
nucleo-chronometry becomes possible based on the abundance of e.g., Th and
U. Bright metal-poor stars are particularly useful for this task since the
weak uranium line at 3859\,{\AA} is only detectable in very high-quality
spectra. The availability of U, together with Th, as the U/Th chronometer
drastically reduces the theoretical uncertainties on the derived ages, as
compared to the more easily detectable Th/Eu ratio \citep{wanajo2002}. Hence,
a lower limit for the age of the Universe can be derived independently of
other measurements such as WMAP \citep{WMAP}. Many \mbox{C-enhanced}
metal-poor objects (hereafter CEMP; $\mbox{[C/Fe]}\footnote{We use the common
notation of $\mbox{[A/B]} = \log({N_{\rm A}/N_{\rm B}})_{\star} -\log({N_{\rm
A}/N_{\rm A}})_{\odot}$, for elements A and B.}>1.0$) were discovered as they
occur with increased frequency amongst metal-poor objects (e.g.,
\citealt{1999rossicarbon}). One of those is the very iron-deficient giant
HE~0107$-$5240 ($\mbox{[Fe/H]} =-5.2$)\footnote{Employing the same Fe NLTe
correction as used for HE~1327$-$2326} that was found in a sample of
$\sim2000$ HES metal-poor candidates \citep{HE0107_Nature}. This star provided
critical observational material to the theories of the formation of the first
objects in the Universe.

For elemental abundance studies involving isotopes, such as $^{6}$Li in
metal-poor turnoff stars, or very weak and/or blended lines, such as uranium
in some strongly r-process enhanced objects, it is crucial to be able to
obtain high-resolution ($R>40,000$) spectra with a very high signal-to-noise
ratio. Bright metal-poor stars are thus ideal targets. With the increased
light-collecting power of current 8-10\,m telescopes it is possible to achieve
a $S/N$ of $200$ or more for such objects within reasonable exposure times.

Apart from the faint stars, the HES also contains large numbers of bright
($B<14$) stars. However, those have previously not been investigated. The main
technical reason is the partial saturatation of the photographic plates,
resulting in partially satured spectra. Despite an increased interest to
search for more fainter metal-poor stars in the extended survey volume of the
HES, recovering those brighter stars is well worth the effort, because they
offer a variety of advantages over fainter stars.  In general, for bright
stars, the availability of various data sources, such as the photometric
survey 2MASS \citep{2MASS}, or astrometric catalogs, e.g., US Naval
Observatory CCD Astrograph Catalog (UCAC2; \citealt{UCAC}), or the Southern
Proper Motion Catalog (SPM3.1; \citealt{SPM}), provide useful additional
information for the analysis of large samples of objects.  Hence, more
detailed analyses, including kinematic studies that make use of the full space
motions, of many bright stars can be carried out more quickly and easily.

The HES thus provides the opportunity to quickly identify large numbers of
bright metal-poor stars in a systematic fashion. Such objects provide
important information in the intermediate magnitude range between the early
objective prism surveys and the HES. For example, a second star,
HE~1327$-$2326, with an extremely low iron abundance of $\mbox{[Fe/H]} =-5.4$
\citep{HE1327_Nature, o_he1327,Aokihe1327} was already recently discovered in
the sample of bright HES stars discussed in this paper.

Here we present the sample of bright metal-poor candidates selected from the
HES. The candidate selection is described in \S~\ref{hes}, while the results
of the follow-up observations of the sample are presented in
\S~\ref{followup}. In \S~\ref{ana} we outline our data analysis and the
abundance results of the metal-poor stars found in the sample. In
\S~\ref{concl} we briefly conclude with an outlook for future work concerning
the newly-discovered stars.

\section{ The Hamburg/ESO Survey} \label{hes} 
\subsection{Recovering the bright metal-poor stars}
The HES is an objective-prism survey initially designed to search for bright
quasars ($14<B<18$) in the southern sky \citep{hespaperIII}. Exploitation of
the stellar content amongst the $\sim4$\, million digital spectra was
initiated later. Thus far, 329 (out of 380) HES plates have been searched for
metal-poor stars and other stellar objects, e.g., field horizontal branch
stars \citep{fhb} or DA white dwarfs \citep{stellar_content_I}. The remaining
plates have recently been processed and follow-up work is now underway.

The extraction and selection of objects in the HES from the photographic
survey plates is described in the following. After the construction of an
input source catalog using the Digitized Sky Survey (DSS), an astrometric
transformation was established between the direct (photographic) and spectral
plates that yielded the position of each object on the spectral plate. Thus,
wavelength zero points were derived. The wavelength range of all HES spectra
is set by the atmospheric cut-off at $3200$\,{\AA} and the sharp sensitivity
cut-off of the IIIa-J emulsion at $5400$\,{\AA}. The survey spectra in the HES
database were first classified into three groups: \texttt{stars, ext, bright},
with \texttt{stars} standing for spectra of point sources, \texttt{ext} for
spectra of extended objects, and \texttt{bright} for possibly saturated
spectra of point sources.  These classes of spectra were extracted with
different algorithms \citep{hespaperIII}. In particular, the spectra of
objects above a saturation threshold were extracted by summing the values of
the photographic density perpendicular to the dispersion direction in a wide
range, and hence incorporating unsaturated regions of the cross-dispersion
profile. This results in an extension of the dynamic range of the
objective-prism plates by 2-3 magnitudes.  Each object was assigned a 12
character HES designation such as ``HE~1234$-$5678''. The digit combination is
based on the B1950 coordinates of the object.

\begin{figure}
\includegraphics[width=7cm,clip=true]{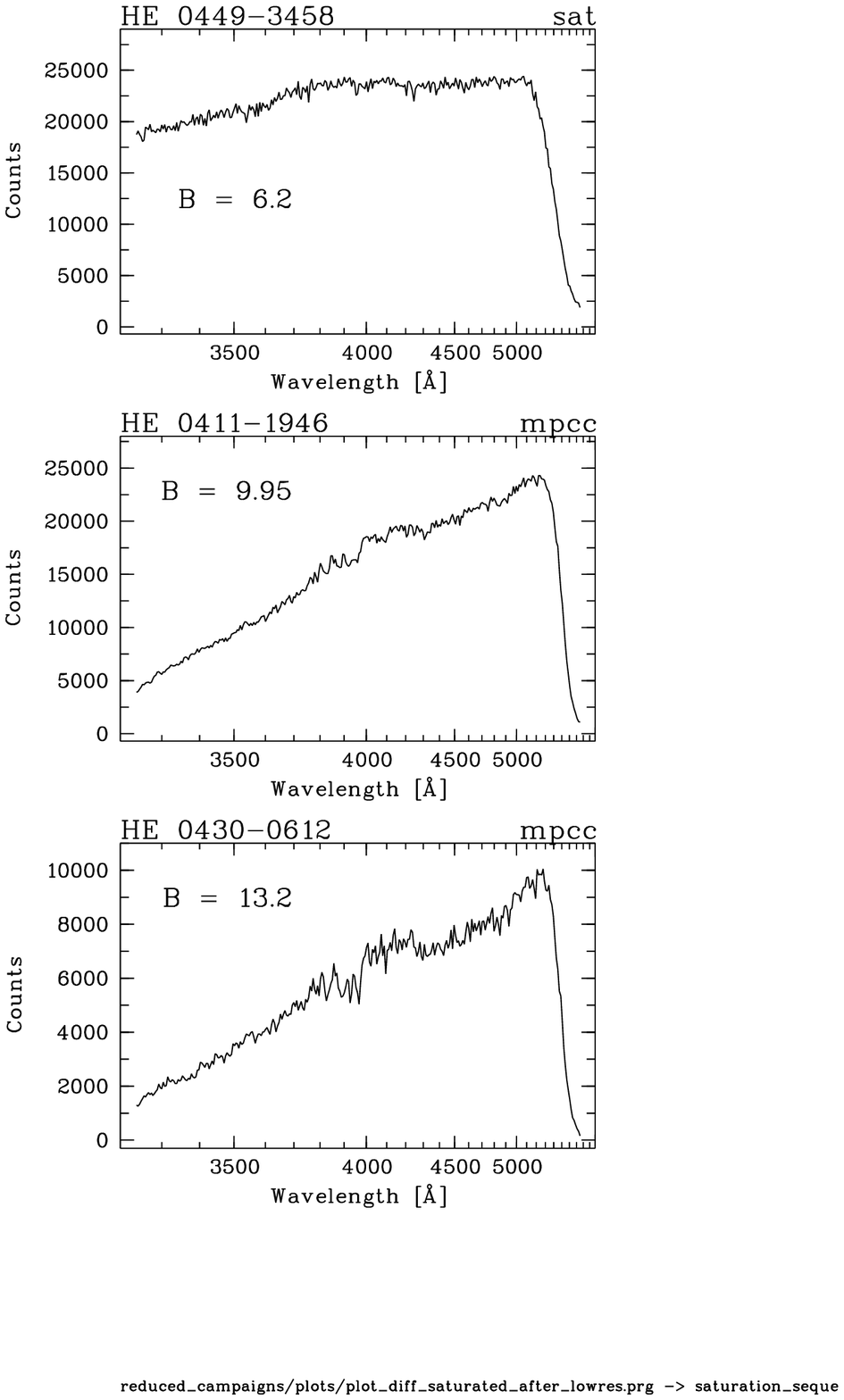}
\figcaption{\label{diff_groups} Spectra displaying a sequence of saturation
  levels occurring in the survey. For comparison purposes, the three stars
  have been chosen to have similar color ($B-V\sim 0.45$). Saturation of the
  emulsion depends on the brightness of the object and the individual
  photographic plate. Typically, from $\sim10,000$ counts onwards, the spectra
  are formed in the non-linear part of the characteristic curve of the
  photographic emulsion (which slightly varies with each plate). We call these
  spectra ``saturated''. However, as shown in this figure the level of
  saturation varies and hence the recoverability of the data. The spectrum in
  the upper panel has been chosen in the visual selection as completely
  saturated (\texttt{sat}) and was excluded from further processing (for the
  definitions of the different classes see the text). The middle panel shows a
  spectrum which has been regarded as recoverable and thus selected as
  \texttt{mpcc} (see \S~\ref{vis}). For comparison, an unsaturated spectrum is
  shown in the lower panel.}
\end{figure}

This paper investigates the stellar HES spectra classified as \texttt{bright}
only. The main difference between \texttt{star} and \texttt{bright} is a
brightness cut-off defined by the level of saturation.
Figure~\ref{diff_groups} shows three \texttt{bright} survey spectra to
demonstrate the different levels of saturation.

\subsection{The calibration}\label{cal}

We seek to select metal-poor stars in a wide color range, i.e., stars near the
main-sequence turnoff point as well as subgiants and giants.  Thus, it is
important to apply a selection that takes into account the strength of the
{\cak} line (which is the strongest metal absorption line in the optical
wavelength range) as a function not only of the metallicity of a star, but
also of its effective temperature. The strength of the {\cak} line can be
measured in the HES spectra by means of the line index \texttt{KP} (a
pseudo-equivalent width measurement in {\AA }) as defined by
\citet{BeersCaKII}.

\begin{figure}
 \begin{center}
\includegraphics[clip=true]{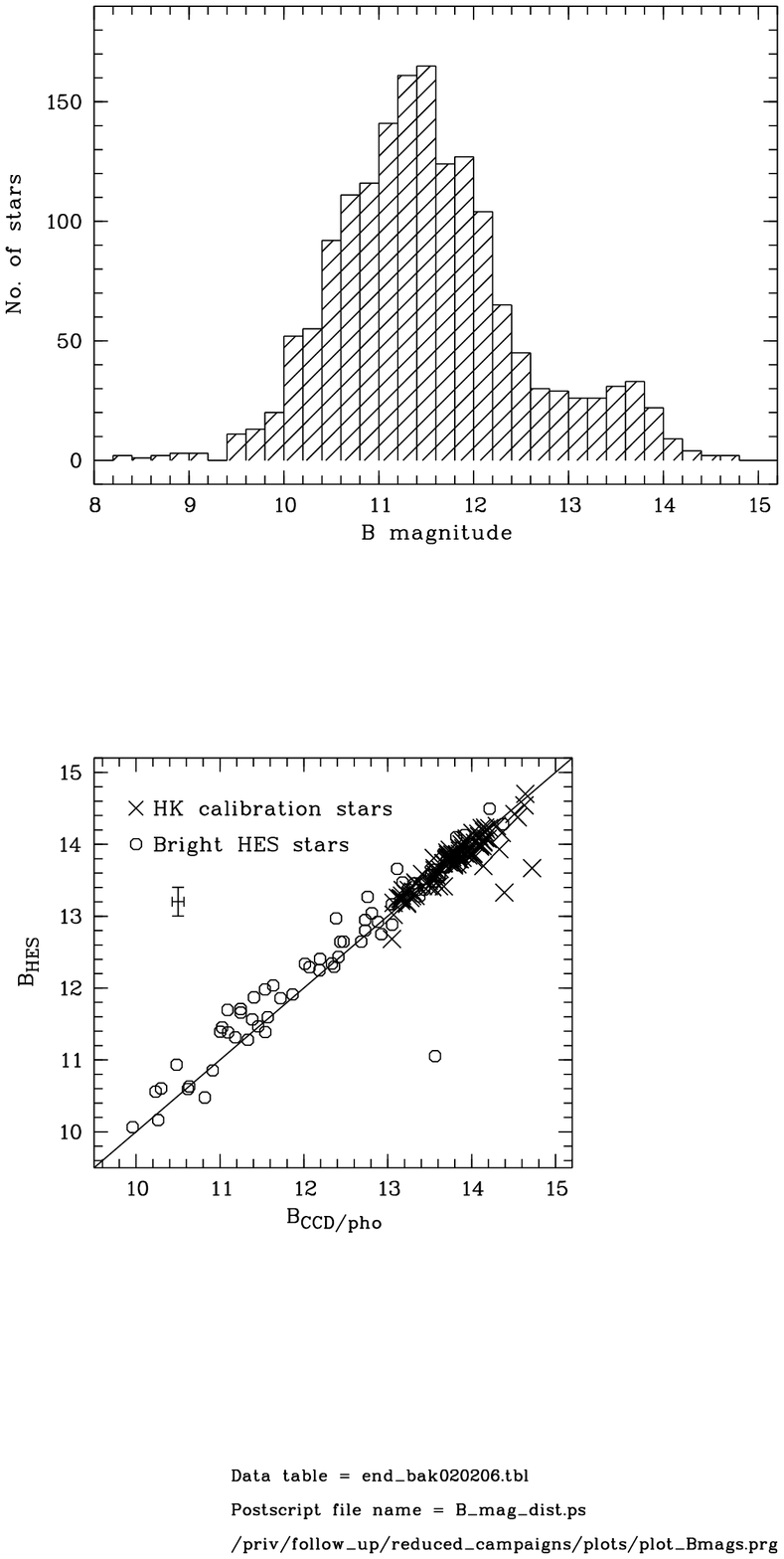}
\figcaption{\label{Bmag} Comparison of the $B$ magnitude for the 85 stars in
  the sample (\textit{open circles}) with available CCD photometry and the 298
  HK calibration stars with photoelectric photometry (\textit{crosses}). A
  typical error bar is also shown. As can be seen, the HK calibration sample
  is fainter than $B\sim13$ and the agreement with the HES magnitudes is
  suitable for the search for metal-poor stars. However, no information could
  be gained concerning the magnitude quality of the brighter stars. This was
  only achieved \textit{after} medium-resolution follow-up observations. The
  additional photometric data of brighter stars revealed that the HES
  magnitude measurements are somewhat influenced by the saturation
  effects. The accuracy of the magnitudes was considered sufficient for our
  current work.}
 \end{center}
\end{figure}

\begin{figure}
 \begin{center}
  \includegraphics[clip=true]{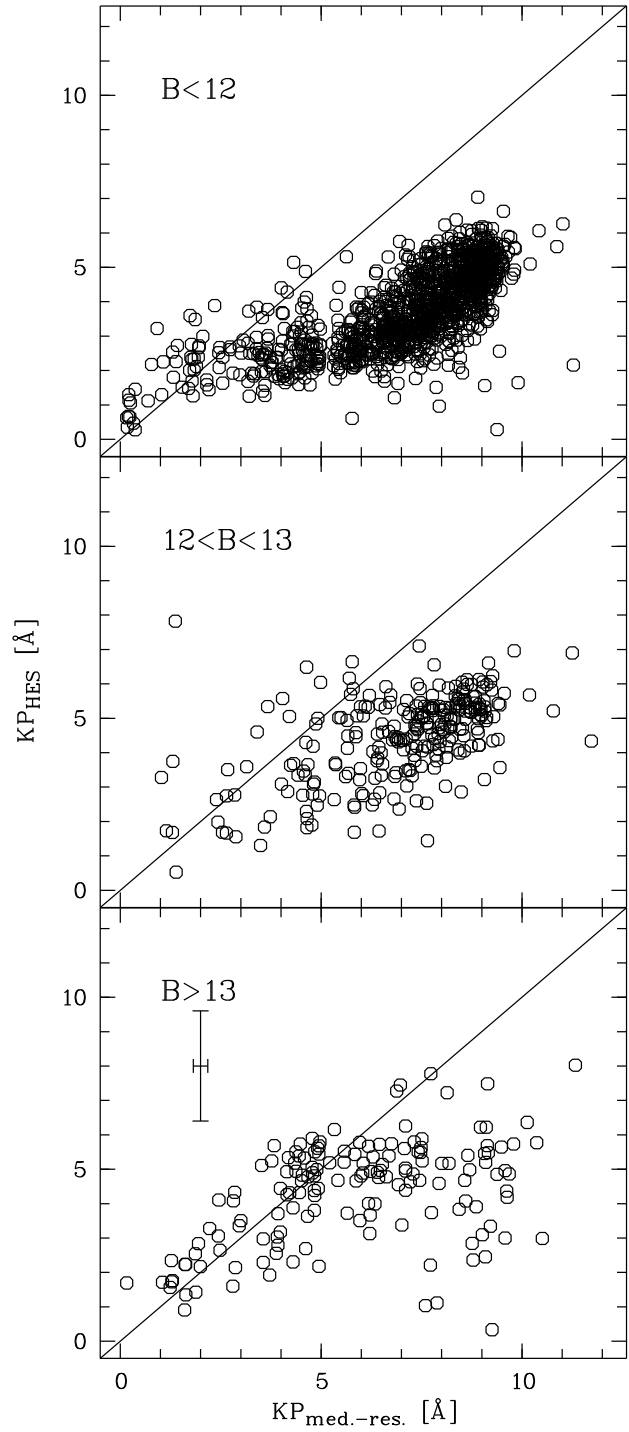}
\figcaption{\label{kp_comp}Comparison of the \texttt{KP} index for all sample
  stars on the photographic plates and in the medium-resolution follow-up
  spectra. The measurement accuracy of the survey index is clearly affected by
  saturation of the spectra of stars with $B<13$. A typical error bar is shown
  in the bottom panel. See text for further details.}
 \end{center}
\end{figure}

The algorithm we employ here to select such metal-poor candidates had
previously only been used in conjunction with unsaturated spectra of faint
stars (i.e., \texttt{stars}). Hence, it was first necessary to test whether or
not the selection algorithm would be applicable to partially saturated
spectra.  We verified with HK survey stars \citep{BPSII} present on HES plates
that the HES \texttt{KP} measurements are in good agreement with the
measurements made in moderate-resolution (i.e., $\Delta\lambda \sim 2$\,{\AA})
follow-up spectra. However, the 298 HK survey stars used for this verification
cover the magnitude range $13 < B < 14.5$, which corresponds only to the faint
end of the bright sample (see Figure~\ref{Bmag} for the $B$ distributions of
the two samples). Later, we learned that for the brighter stars the saturation
of their spectra becomes stronger, and results in a systematic underestimate
of the \texttt{KP} index. Unfortunately, the same problem occurs for any stars
with $\texttt{KP}>5$\,{\AA}, where the survey data underestimates the index
strength, regardless the brightness of the star. It follows that the selection
efficiency of metal-poor giants is low, because more metal-rich giants were
mistaken as metal-poor. However, no metal-poor stars should be lost due this
problem. This is illustrated in Figure~\ref{kp_comp}, where the \texttt{KP}
measurements from the prism-survey are compared with those obtained from
medium-resolution spectroscopic data.

\begin{figure}
 \begin{center}
  \includegraphics[clip=true]{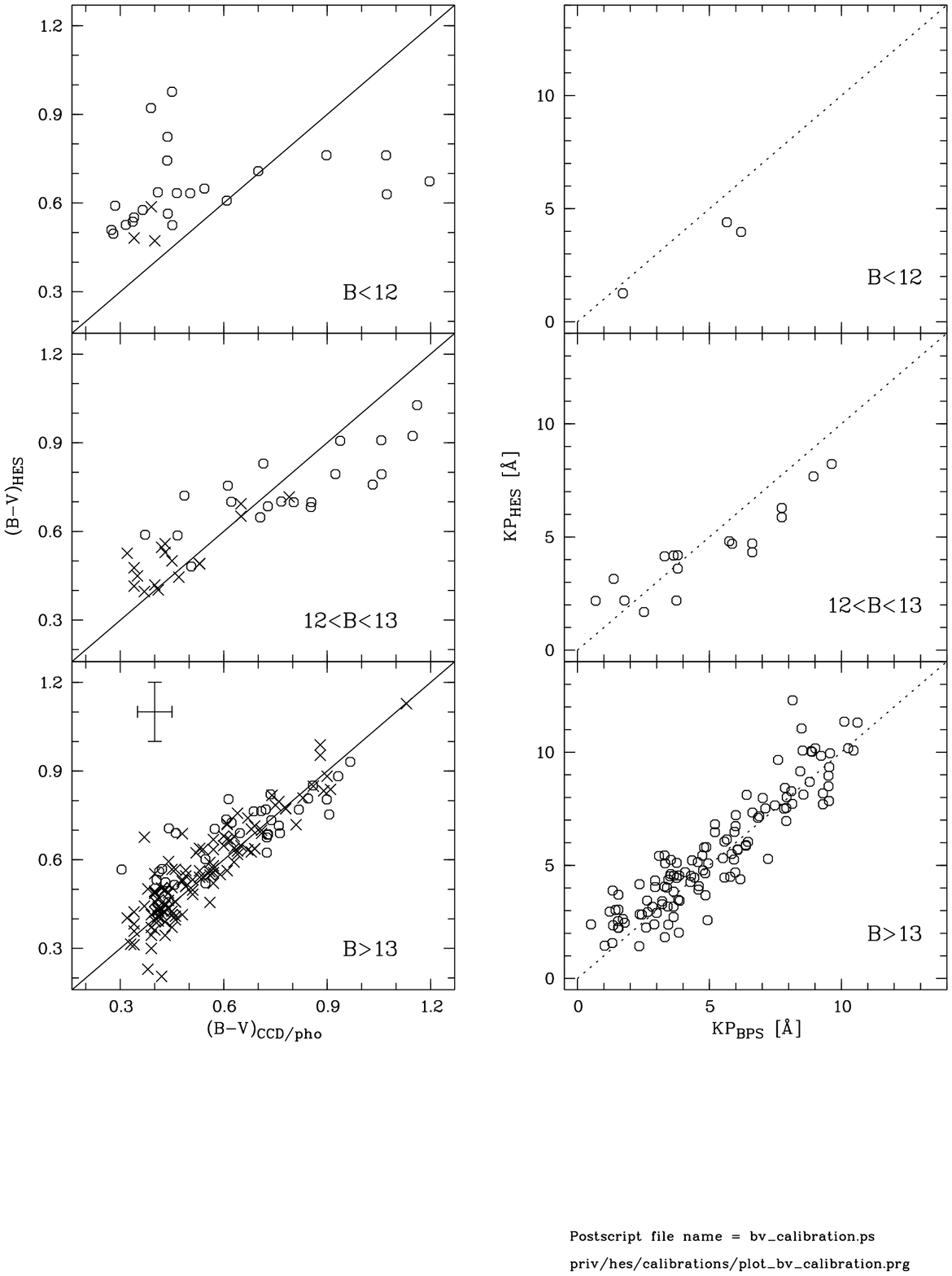}
\figcaption{\label{bv_calib} Comparison of CCD/photoelectric and survey $B-V$
   colors as a function of brightness. Symbols are the same as in
   Figure~\ref{Bmag}. The scatter becomes much larger for stars brighter that
   $B=13$, which reflects the onset of saturation effects in the spectra of
   these objects. A typical error bar is shown for the fainter stars.}
\end{center}
\end{figure}

It is also possible to measure the $B-V$ color directly from the HES spectra,
using the ``half power point'' technique described by \citet{hespaperIII}.
The $B-V$ color of the unsaturated HES objects has an uncertainty of 0.1\,mag
\citep{stellar_content_II}. Figure~\ref{bv_calib} shows that equally good
results can be achieved for the HES \texttt{bright} stars fainter than
$B\sim13$. However, for the brighter stars the scatter increases
significantly, and the $B-V$ color of the redder objects is systematically
underestimated. This can be explained by the fact that the photographic
emulsion is more sensitive in the red part of the HES spectra than it is in
the blue. Thus, the saturation effects should appear first in the red, and
they should be more pronounced for red stars. The consequences of this issue
for our sample are further described in \S~\ref{ccd}.

\subsection{\label{sel}The selection}
In this section we briefly describe the selection of metal-poor candidates
amongst \texttt{bright} HES stars. Initially, no CCD photometry was available
for HES stars brighter than $B\sim13$. Based on the results from the
comparison with the HK survey sample, we concluded that the overall level of
saturation was not too strong, and that the quality of our survey spectra
would be sufficient to begin the search for bright metal-poor stars.

In the first step, we restricted the range of objects to be considered to
$0.3<B-V<1.2$. This corresponds to stars redder than the main-sequence turnoff
point for metal-poor stars of an age of about 12\,Gyr, and bluer than the
stars at the tip of the red giant branch. On both sides of the color range,
the chosen cut-off allows for an error margin of about 0.1\,mag. During the
selection, we ignored any potential reddening of the stars. The HES stars are
located at high galactic latitudes, and any reddening usually is considerably
smaller than 0.1\,mag (see e.g., \citealt{fhb}).

For the second step of the selection, we determined a cut-off line in the
\texttt{KP} index versus $B-V$ color parameter space in the following way. A
set of simulated stars with \texttt{KP} and $B-V$, equally distributed in the
range 0--15\,{\AA} and 0.3--1.2\,mag, was created. For these simulated stars,
we estimated [Fe/H] using the techniques of \citet{BeersCaKII}. All stars with
metallicities in the range of $-2.6<\mbox{[Fe/H]}<-2.4$ were selected.  A
cut-off line was fit to this set of data. The result will be discussed in
further detail in a forthcoming paper (N. Christlieb et al. 2006, in
preparation). All HES stars with a \texttt{KP} index lower than the cut-off
value determined for its $B-V$ color were selected as metal-poor
candidates. This results in a metallicity cut-off at
$\mbox{[Fe/H]}=-2.5$. Applying this technique yielded $5081$ \textit{raw}
candidates.

In the final step of the selection, the raw and extracted HES spectra of these
objects and the corresponding area of the direct plates available in the DSS
were inspected in order to eliminate false positives due to plate artifacts,
scratches, dust on the HES plates, emulsion flaws, ghosts and object
extraction at a slightly wrong position on the spectral plate (a
mis-positioned extraction results in systematically too low values of the
{\cak} index).

\subsection{The visual inspection}\label{vis}
The apparent strength of the \cak\ line at 3933\,{\AA} was visually estimated
against the continuum. Depending on how strong the line appeared, the objects
were classified into several metal-poor classes (mpc) : \texttt{mpca} --- the
spectrum clearly shows no \cak\ line; \texttt{unid} --- it is unclear if a
\cak\ line is visible in the spectrum or not; \texttt{mpcb} --- a weak \cak\
line is visible; and \texttt{mpcc} --- a significant \cak\ line appears
against the continuum. The remaining stars were grouped into \texttt{star} ---
ordinary (metal-rich) star; \texttt{hbab} --- hot horizontal branch A or B
stars displaying (very) strong Balmer lines; \texttt{art} --- plate artefact;
\texttt{ovl} --- overlap of spectra of optically close stars due to
unfortunate dispersion direction; and \texttt{gal} --- galaxy, the object
appeared to be extended on the direct image of the DSS. We identified 3309
spectra as belonging to the non metal-poor classes.

\begin{figure*}
 \begin{center}
  \includegraphics[width=13cm,clip=true]{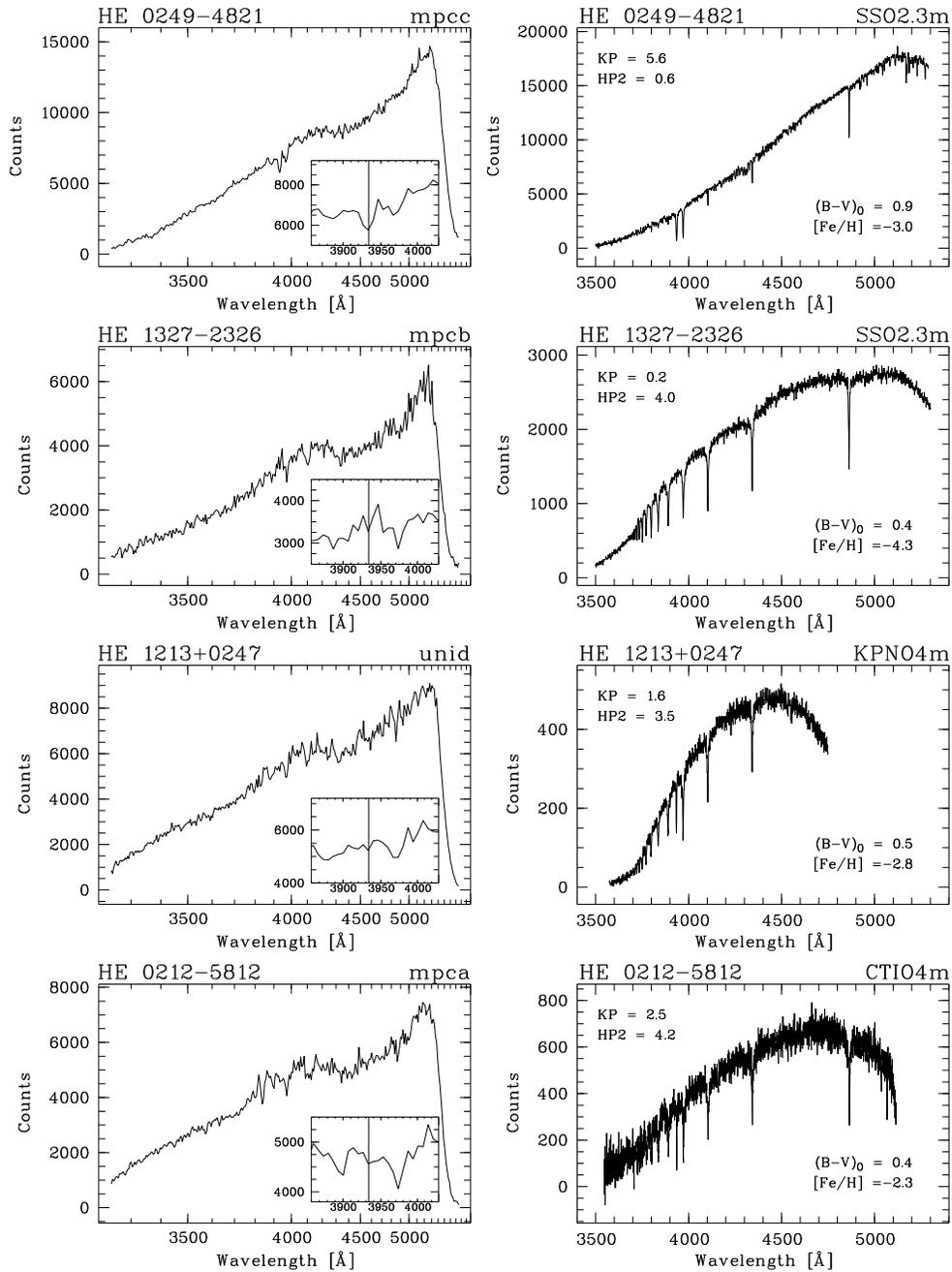}
\figcaption{\label{mpc} Examples of objective-prism spectra (left panel) of the
  four different classes of metal-poor candidates and their corresponding
  medium-resolution follow-up spectra (right panel). From the top: classes
  \texttt{mpcc}, \texttt{mpcb}, \texttt{unid} and \texttt{mpca}. In the insets
  we show an expansion of the \cak\ region by which the survey spectra were
  classified. For further description of the classes, see text. We list the
  telescope employed at the top right corner. The \texttt{KP} and \texttt{HP2}
  index, the $(B-V)_{0}$ color and the final iron abundance of each star are
  shown next to the follow-up spectra (see \S~\ref{data_proc} for more details
  on the indices). Despite the classification, there is no strong correlation
  between iron abundance and class.}
 \end{center}
\end{figure*}

This leaves 1767 bright metal-poor candidates to form our working
sample. Figure~\ref{mpc} shows some examples of the four metal-poor classes,
together with their follow-up spectra (see \S~\ref{followup} for more
details). The division of the candidates into the classes resulting from the
visual inspection is presented in Table~\ref{selection_results}. 

\begin{figure*}
 \begin{center}
\includegraphics[clip=true]{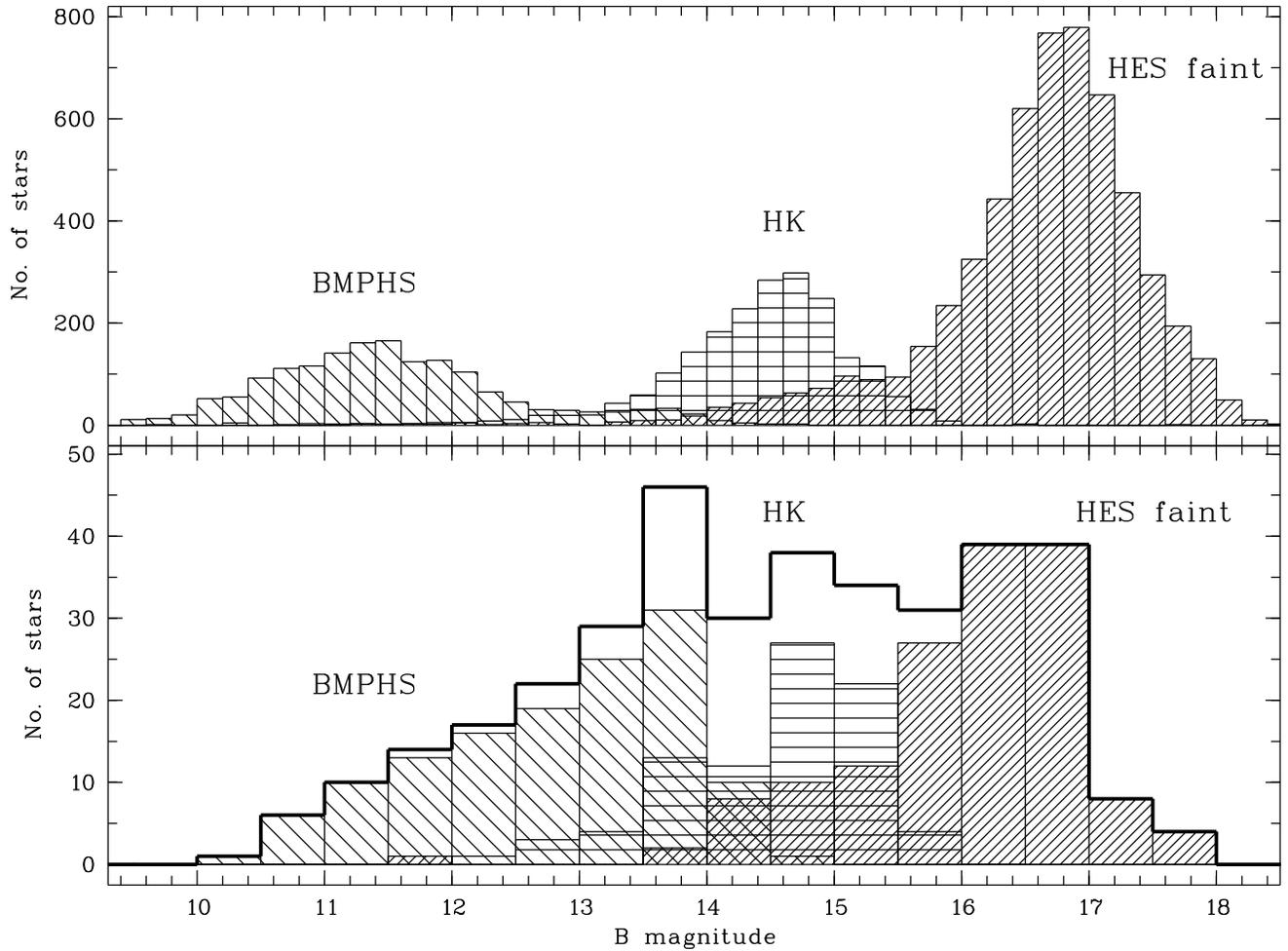}
\figcaption{\label{Bmags} $B$ magnitude distribution for the bright HES stars,
  the HK survey stars and the faint HES stars. The magnitude ranges of the
  samples complement each other and span a large range, from $B\sim 9$ to
  $B\sim 18$. \textit{Top panel}: Note that the bimodal distribution of the
  bright and faint HES stars is due to a selection bias in the sample of
  bright HES stars (see text for details). \textit{Bottom panel}: Same
  distribution as above, but the displayed stars are confirmed to have
  $\mbox{[Fe/H]}<-2.0$ by follow-up spectroscopy, and randomly selected such
  that the total numbers of HK survey stars, bright HES and faint HES stars
  are the same (i.e., 150). Note the smooth distribution, and the removal of
  the strong selection bias towards bright stars in the sample of bright HES
  stars that can be seen in the top panel.}
 \end{center}
\end{figure*}

For 1626 of the total 1767 stars there are HES $B$ magnitudes available; the
remaining stars are beyond the bright end of the magnitude range covered by
the photometric sequences used to calibrate the HES photometry (see
\citealt{hespaperIII} for a description of the HES photometry). The average
magnitude of the bright stars is $B \sim 11.5$\,mag. The B magnitude
distribution of the sample of bright stars is shown in the top panel of
Figure~\ref{Bmags}, together with the distributions of the HK survey and the
faint HES stars. All three samples complement each other in brightness,
covering a range of almost 10 magnitudes.

Concerning the HES, one might expect that the sample of bright HES stars would
smoothly add objects to the bright tail of the distribution of the faint HES
stars. However, this is not the case. Rather, the combined HES stars show a
bimodal distribution. The form of the distribution can be explained by a
selection bias present in the sample of bright metal-poor candidates from the
HES. As discussed in \S~\ref{sel} above, the saturation of the HES spectra of
bright stars results in a systematic underestimation of their {\cak} line
index, hence a large number of false positives enter the candidate
sample. This effect is strongest for the brightest stars, thus, bright stars
are strongly over-represented in the sample of bright metal-poor candidates.

If this explanation is correct, than the bimodal distribution should not be
seen in the sample of \textit{confirmed} HES metal-poor stars. Indeed, as
shown in Figure \ref{Bmags} (thick lined envelope of bottom panel), the
magnitude distribution is smooth when the false positives are removed.

\subsection{Cross-correlation with other catalogs}\label{cross}

In order to identify re-discovered stars in our sample and to gain additional
information for as many sample stars are possible, a cross-correlation of the
HES database with several catalogs was carried out. We identified stars in the
2MASS All Sky Data Release \citep{2MASS} and obtained $J$, $H$ and $K$
magnitudes, in the UCAC2 \citep{UCAC} and in the SPM3.1 \citep{SPM} catalog to
retrieve proper motions. We then cross-correlated with the HK survey database,
and found a number of stars from the HK survey in our sample.  We also made
use of the following catalogs: the non-kinematically selected stellar sample
from \citet{Bidelman_MacConnell}, the kinematically selected metal-poor
stars samples of \citet{Ryanetal:1991}, and \citet{carney_cat}. Finally, we
made extensive use of the SIMBAD database. During this search many of our
stars appeared in well-known surveys, such as the Guide Star Catalog, the
Positions and Proper Motion Catalog, or the HIPPARCOS/TYCHO catalog. See Table
\ref{num_id} for the numbers of stars found in the catalogs and the SIMBAD
database\footnote{We generally used an 8'' search radius.}.

We are aware of $29$ stars in our sample known to have
$\mbox{[Fe/H]}\lesssim-2.0$.  It should be noted though, that, due to our
selection criteria, our sample is supposed to only contain objects with
$\mbox{[Fe/H]}<-2.5$. Hence, it is expected that stars with
$\mbox{[Fe/H]}>-2.5$ are missing from our compilation of rediscovered objects.
Many of these objects were found in several catalogs or studies. Twenty four
of those are metal-poor stars. For $16$ of them, an analysis based on
high-resolution spectroscopy can be found in the literature. The re-discovered
HES stars are listed in Table~\ref{rediscoveries}, together with selected
other names, as well as, where available, iron abundances based on medium-
and/or high-resolution studies. For comparison, in column (2) we also list our
final iron abundance estimate.  Amongst the re-discovered stars are G64-12
(HE~1337+0012; \citealt{1981G64-12}) and the well-known strongly r-process
enhanced metal-poor star \mbox{CS~22892-052} (HE~2214$-$1654;
\citealt{Snedenetal:1996}). These re-discoveries show that our technique is
successful in finding metal-poor stars. We flagged the re-discovered stars as
such and kept them in our database as reference objects and for statistical
purposes (see \S~\ref{MDF}).

In Table~\ref{other_redisc} we list five rediscovered stars for which we
determined $\mbox{[Fe/H]}<-2.0$, but which are horizontal-branch stars.  From
these rediscoveries, we learn that our technique also works for other types of
stars (as long as the indices and colors are within their acceptable ranges):
HE~0411$-$3558, for example, is a horizontal-branch (HB) star discovered by
\citet{BPSII} (CS~22186-005). They measured $\mbox{[Fe/H]}=-2.66$ from a
medium-resolution spectrum, while \citet{ryan96} obtained
$\mbox{[Fe/H]}=-2.77$ from high-resolution data. Our final estimate is
$\mbox{[Fe/H]}=-2.86$, which agrees well with the two other iron abundances.

Our follow-up spectra do not include any gravity-sensitive indicators
available which is accurate enough for allowing the determination of
evolutionary status of the stars. Hence, our sample of metal-poor stars is
likely to have some contamination by HB stars. The rediscovery of a HB star
and a red HB star in our sample (as listed in Table~\ref{other_redisc}) for
which we measured $\mbox{[Fe/H]}<-2.0$ confirms this. To quantify the expected
level of contamination in our sample, we determined the fraction of HB stars
to metal-poor star in the catalog of \citet{BPSII}. Based on gravity-sensitive
photometry, they classified all their spectra according to evolutionary
status. It appears that amongst the more metal-poor stars in their sample, the
frequency of HB stars is $5\%-10\%$. They also find $\sim2\%$ RR Lyr
variables. We thus conclude that the contamination by HB stars is acceptably
low for our work, and does not pose any concerns to our conclusions.

Trying to identify the known metal-poor stars in our sample also suggested
searching for stars that may have been lost in the process of visual
inspection of the raw candidates. We checked all stars discarded during the
inspection in the SIMBAD database. Of the 3309 stars, $\sim1100$ had some sort
of entry, and $\sim$250 of these had at least one reference listed. Only 5 
stars have $\mbox{[Fe/H]}<-2.5$\footnote{A further 27 objects were later found
to have been classified as metal-poor stars, but due to an unfortunate
incident they were lost when producing the final list of 1767 visually
inspected metal-poor candidates. Amongst those is the well-known metal-poor
giant CD $-$38~245 \citep{cd38}. These objects will be included in our ongoing
studies of the remaining 51 HES fields.}.
Table~\ref{non_mp_resdisc} gives further details, including our classification
of these stars. The two \texttt{hbab} classifications are not in disagreement
with the original classification found in the literature (i.e., they are a hot
turnoff star and horizintal branch star). Also, the object classified as
\texttt{ovl} could technically not have been picked up as
metal-poor.

From a cross correlation with the HK survey database, we identified 44 stars,
of which 30 are known to be A-type stars \citep{wilhelm99, beers96_fhb}. Of
those 30, we classified 27 as \texttt{hbab}, while the remainder are
\texttt{ovl}. The 4 emission-line objects and the one RR Lyr star were all
correctly classified as \texttt{star}, and \texttt{hbab}, respectively. The
remaining HK stars are more metal-rich than $\mbox{[Fe/H]}=-2.5$. 

We conclude that no stars were missed that should have been selected by our
algorithm. Based on this result, we infer that our visual selection mechanism
works very well to select the most metal-poor stars in a given sample of
pre-selected HES candidates.

\section{Medium-resolution follow-up spectroscopy}\label{followup}
\subsection{Observations}
The HES objective-prism spectra have a resolution of $\sim 10$\,{\AA} (i.e.,
$R=\lambda/\Delta\lambda\sim400$) at the \cak\ line, and an average S/N ratio
of $\sim 8/1$, which is of sufficient quality to select metal-poor candidates
in large numbers. To obtain a more reliable estimate of the metallicity than
was known from the survey data, higher-quality spectra were needed.  Hence,
medium-resolution follow-up observations were taken with several telescopes
(SSO $2.3\,$m, KPNO $4\,$m, CTIO $1.5\,$m, ESO $3.6\,$m, CTIO $4\,$m,
AAT\footnote{Obtained during cloudy weather in a program not related to
metal-poor stars of Ryan et al.} $3.9$\,m) in the periods Mar -- Apr 2003, Aug
2003 -- Apr 2004, and Sep -- Oct 2004. The spectra were obtained with a
wavelength range covering at least $3600-4800$\,{\AA}, with $R \sim 2,000$ and
$S/N > 20/1$ at the \cak\ line. Exposure times during good weather conditions
varied between $t_{\rm {exp}}\sim 20-300$\,s. Using longer exposure times,
these stars could also be observed through thin clouds or during periods with
poor seeing (i.e., FWHM~$>2\farcs0$). See Table~\ref{med-res} for more
details on the follow-up observations. Due to the low spatial resolution of
the DSS, from which the HES input catalog was created, several spectra of two
stars with small angular separation (i.e., $\lesssim 3\farcs0$) were not
spatially resolved on the survey plates. Hence, the ``pair'' was recorded with
only one HES designation. We identified $10$ of these ``pairs'' in the course
of the medium-resolution observations. Since it was not possible to decide
which of the two stars would have been the correct metal-poor candidate, both
were observed, and the sample size consequently increased to 1777 objects.

\subsection{Data reduction and processing} \label{data_proc}

Depending on where the data were obtained, the reduction of the spectra was
carried out using different software. SSO data were fully reduced with Figaro
\citep{figaro} and Fortran programs, the KPNO and CTIO data with
IRAF\footnote{IRAF is distributed by the National Optical Astronomy
Observatories, which are operated by the Association of Universities for
Research in Astronomy, Inc., under cooperative agreement with the National
Science Foundation.} routines, the ESO data with ESO/MIDAS \citep{midas}, and
the AAT data with IRAF.

Radial velocities of the targets were determined, and the spectra shifted to
the restframe. To obtain the most accurate radial velocities as possible from
medium-resolution spectra we used the combined results of two different
methods: (1) Measuring the position of the Balmer lines H$\beta$, H$\gamma$
and H$\delta$ and (2) Fourier cross-correlation template matching with a grid
of model atmosphere synthetic spectra with different temperatures, gravities
and metallicities.

From comparing the results obtained with the two methods, we found a
systematic offset of $\sim10$\,km\,s$^{-1}$ of the cross-correlation method
with respect to the Balmer line results. However, we verified with
observations of several standard stars taken during the program that the
Balmer line method had no offset. To correct this effect, we removed the
offset from the cross-correlation radial velocities. We adopted a weighted
average of the two radial velocity measurements, where available.  The
velocity uncertainties of the two methods are of the same order.  The median
of the weighted uncertainty in the Balmer line method is $\sim7$\,km\,s$^{-1}$
whereas it is $\sim4$\,km\,s$^{-1}$ in the cross-correlation method.  The
advantage of the cross-correlation method is that information from the entire
spectrum is used rather than just three individual lines as in the Balmer line
method. This leads to a more precise velocity determination with a lower
uncertainty.  The median of final, weighted velocity uncertainty is
$\sim3$\,km\,s$^{-1}$.  Further systematic uncertainties, e.g., due to
telescope/instrument instabilities, have not been considered.

We seek to use $\mbox{[Fe/H]}$ as a metallicity indicator of the
stars. However, iron lines are not detectable in our medium-resolution
spectra, particularly in those of our targeted metal-weak stars. Instead, the
\cak\ resonance line at $3933$\,{\AA} is very strong and can easily be
measured. We employ the \cak\ calibration of \citet{BeersCaKII} to estimate
$\mbox{[Fe/H]}$ for each program star. This is done using the \cak\ line index
\texttt{KP} \citep{BeersCaKII} and $B-V$. The \texttt{KP} index is obtained by
comparing the mean flux of two continuum regions around the \cak\ line with
the flux of the \cak\ line itself. Regarding the color, we use two differently
derived estimates. (1) The color $(B-V)_{\rm J-K}$ obtained from the
dereddened 2MASS $J-K$ (see \S~\ref{ccd}), and (2) the spectroscopically
derived $(B-V)_{\rm HP2}$ color based on the \texttt{HP2} line index and the
\citet{BeersCaKII} method. The \texttt{HP2} index is a measurement of the
strength of the H$\delta$ line, which depends on the effective temperature of
the star, and hence its color. An advantage of using a color derived from the
\texttt{HP2} index is its independence of reddening.  Finally, we measure the
Beers et al. \texttt{GP} index from a comparison of the mean flux of two
continuum regions with the flux of the CH band (G-band) at 4300\,{\AA}. We use
the \texttt{GP} index to calculate carbon abundance estimates of all stars
with $\mbox{[Fe/H]}<-1.0$ (see \S~\ref {carbon_ana}).

\subsection{Comparison with accurate photometry} \label{ccd}
CCD photometry was obtained for 84 stars in the sample\footnote{These data
were taken by the Beers et al. collaboration as part of the ongoing effort to
collect photometry for metal-poor candidates.}. Additional photoelectric
photometry (e.g., \citealt{BPSI,BPSII, norrisUBVpho, Anthony-TwarogUBVpho}) is
available for 298 HK survey stars rediscovered in the HES (see also
\S~\ref{cal}). We compared these data with the HES $B$ magnitudes initially
derived from the DSS direct plates. The agreement is good, as shown in
Figure~\ref{Bmag}. The saturation effects present in the spectra, however,
were expected to have an influence on our $B-V$. We used the combined data to
test this. Figure~\ref{bv_calib} presents the comparison of the $B-V$
colors. There is a clear deviation from the one-to-one $B-V$ relationship with
the brightness of the objects. From a division of the stars into three
brightness bins it is apparent that the saturation effects set in for stars
brighter than $B=13$, thus spoiling the survey color measurements for these
stars. In retrospect, this somewhat contradicts our previous finding that the
survey color measurements would not be strongly influenced by saturation
effects (see \S~\ref{cal}). However, this discrepancy can be understood for
the following reason. Previously, we used a sample that was too faint to be
consistent with our average brightness of $B\sim11.5$\,mag.  

We had no knowledge at the outset of our project that the onset of the
saturation occured at $B\sim13$, which is the brighter limit of the HK survey
calibration sample. As a consequence, we decided not to use the survey colors
but replace them with new $B-V$ derived from the 2MASS $J-K$ colors. We
obtained a regression for a sample of stars with known $B-V$ and $J-K$ colors
(i.e., $B-V = 0.2 + 0.642 (J-K) + 0.707 (J-K)^{2}$). For the cases where no
$J-K$ color was available we used colors derived from the \texttt{HP2} index,
failing which we adopted the original survey color. The new $B-V$ colors agree
well with those derived from the \texttt{HP2} index for stars with
$B-V<0.7$. As expected, for stars cooler than this, the index becomes
insensitive to temperature and the quality of the colors degrades.

\begin{figure}
 \begin{center}
\includegraphics[clip=true]{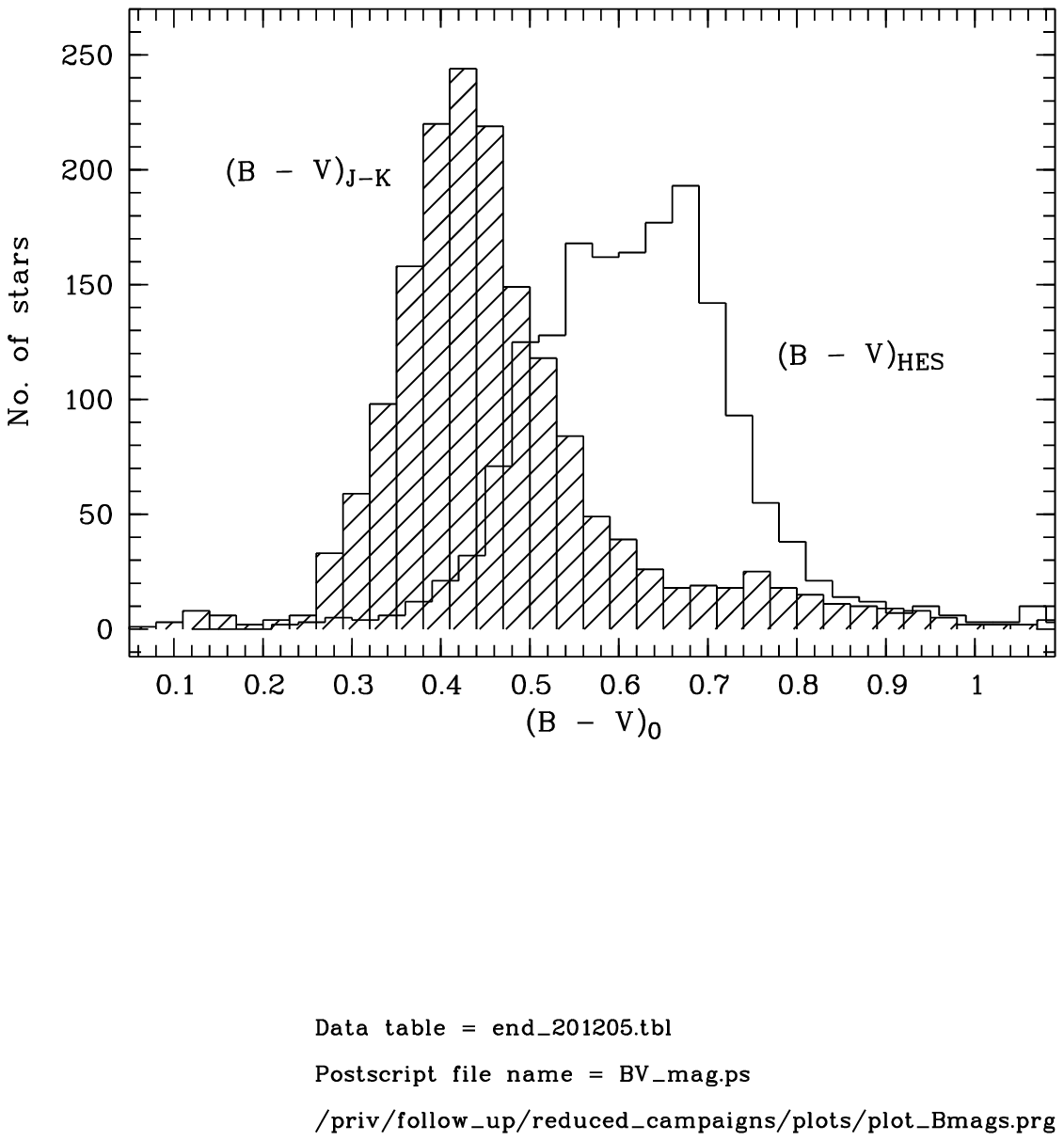}
\figcaption{\label{BVmags}$B-V$ color distribution for the bright metal-poor
   candidate sample as derived from the 2MASS $J-K$ data. When selecting the
   sample from the HES database, a temperature cut-off at $B-V=0.3$ has been
   made, based on the original HES colors. Due to the new colors, a few stars
   now have $B-V<0.3$ which is slightly hotter than has been regarded useful
   for our search. It is interesting to note here that the sample selected on
   the basis of the HES colors was originally thought to contain mostly
   giants. The new $B-V$ colors distribution suggests that the majority of the
   sample stars are dwarfs and field horizontal branch stars. This color shift
   is also reflected in Figure~\ref{bv_calib}.}
 \end{center}
\end{figure}

The $(B-V)_{J-K}$ color distribution is shown in Figure~\ref{BVmags} (hatched
distribution). The majority of stars are turnoff stars ($B-V\sim0.4$), with
several stars hotter than $B-V=0.3$, which was our initial blue limit for the
selection of the sample. From the clear distribution of initial HES colors
in Figure~\ref{BVmags}, it is apparent that the new, $(B-V)_{J-K}$, colors are
bluer than $(B-V)_{\rm {HES}}$, and in agreement with the results from the
photometric colors. This effect could explain why many A-type stars, rather
than metal-poor stars, were found in the sample (for further discussion, see
\S~\ref{gen_fin}). 

Concerning the selection algorithm, we conclude that it will function
correctly only if the survey $B-V$ color is replaced with a better estimate,
such as one based on the 2MASS data.

\section{Analysis of the sample}\label{ana}
\subsection{Iron abundance estimates}\label{feh}

As described in \S~\ref{data_proc}, we use the \cak\ line to estimate the iron
abundance for each star. However, we first excluded all stars from the sample
that were obviously misclassified, e.g., objects with He lines or \cak\
emission-line cores in the medium-resolution spectra. After that, we attempted
to obtain abundances for stars with a \texttt{KP} index, in the range
$0.0<\texttt{KP}\le9.5$ in three ways:

\begin{enumerate}
\item $\mbox{[Fe/H]}_{\rm {B-V}}$: Employing the Beers et al. (1999)
calibration using the \texttt{KP} index together with the color derived from
the dereddened 2MASS $J-K$ data, $(B-V)_{\rm J-K}$. For this method to be
applicable, the color has to lie in the range $0.3<(B-V)_{\rm {J-K}}<1.0$.

\item $\mbox{[Fe/H]}_{\rm {HP2}}$: Employing the Beers et al. calibration
using the \texttt{KP} index, together with the color derived from the
\texttt{HP2}, $(B-V)_{\rm HP2}$. The index has to lie in the range
$0.25\le\texttt{HP2}\le5.3$

\item $\mbox{[Fe/H]}_{\rm {Rossi}}$: Iron abundance estimates using the
regression provided by \citet{rossi_carbon05}, based on the \texttt{KP} index
and the 2MASS $J-K$ color. For the $J-K$ de-reddening, we adopted the approach
of \citet{reddening_bonifacio} to infer $E(J-K)$ from the \citet{schlegel}
$E(B-V)$.  Following Rossi et al., the color has to lie in the range $0.2\le
(J-K)_{0}\le0.8$.

\end{enumerate}

The ranges of the indices and colors are imposed by the limits of the
calibrations used.  Depending on the above criteria, not all methods could
always be employed for each star, and many have only a partial set of iron
abundance estimates.

We note that this work focuses on the identification of the more metal-poor
stars in the sample. We thus concentrate on all objects with
$\mbox{[Fe/H]}<-1.0$, and do not give further consideration to stars with
higher metallicities. $\mbox{[Fe/H]}=-1.0$ is the upper metallicity limit of
all the calibrations employed in this work, because the \texttt{KP} index
begins to saturate for stars more metal-rich than $\mbox{[Fe/H]}\sim-1.5$
\citep{BeersCaKII}. We did not attempt to make use of the auto-correlation
function technique (ACF; \citealt{BeersCaKII}) to determine iron abundance
estimates for the stars more metal-rich than $\mbox{[Fe/H]}\sim-1.5$ in our
sample. This would be beyond the scope of the present investigation of finding
the more metal-poor stars in this sample. However, for a full kinematic
analysis it would be desirable to obtain better abundance estimates for such
stars.

To best determine whether a star has $\mbox{[Fe/H]}>-1.0$ we applied the
following procedure. We weighted and averaged the abundances
$\mbox{[Fe/H]}_{\rm {B-V}}$ and $\mbox{[Fe/H]}_{\rm {HP2}}$ to form the best
estimate based on the \citet{BeersCaKII} method, $\mbox{[Fe/H]}_{\rm
{HP2/B-V}}$. Compared to the values determined from the Rossi et al. 2005
calibration, we regarded the former to be the more robust estimates. We then
discarded the iron abundance estimate of all stars with $\mbox{[Fe/H]}_{\rm
{HP2/B-V}}>-1.0$.  On the other hand, when the star had $\mbox{[Fe/H]}_{\rm
{HP2/B-V}}<-1.0$, we averaged the $\mbox{[Fe/H]}_{\rm {B-V}}$ and
$\mbox{[Fe/H]}_{\rm {Rossi}}$ values to form a best estimate based on the
$J-K$ colors, $\mbox{[Fe/H]}_{\rm {J-K}}$. The $\mbox{[Fe/H]}_{\rm
{HP2/B-V}}$ was not further considered. Finally, we were left with one
abundance derived from the \texttt{HP2} index ($\mbox{[Fe/H]}_{\rm {HP2}}$),
as well as one obtained from the $J-K$ ($\mbox{[Fe/H]}_{\rm {J-K}}$).

Before calculating the final iron abundance from the $\mbox{[Fe/H]}_{\rm
{HP2}}$ and $\mbox{[Fe/H]}_{\rm {J-K}}$ for the remaining stars, we weighted
the methods according to their temperature sensitivity. We tested the change
of the iron abundances to small variations of the input \texttt{HP2} indices
and $J-K$ colors. From that, we estimated that for stars with
$\texttt{HP2}>2.5$ the iron abundances derived from the \texttt{HP2} indices
are more robust than those based on the $J-K$ colors. For stars with
$\texttt{HP2}\le2.5$, both methods appeared to be about equally good.
$\texttt{HP2}=2.5$ corresponds to $B-V\sim0.5$, which automatically results in
a division of dwarfs ($\texttt{HP2}>2.5$) and giants ($\texttt{HP2}\le2.5$).
We note that in a magnitude-limited survey like ours, stars at $B-V>0.5$ are
almost exclusively giants. The survey volume for cool dwarfs is neglible
compared to the survey volume for the higher luminosity giants, which are seen
up to much larger distances. In the following, we refer to dwarfs and giants
according to their $\texttt{HP2}$ index.  This is also the lower, cooler,
limit at which the \texttt{HP2} index begins to lose sensitivity to the $B-V$
color. Additionally, we split the dwarfs and the giants each into
carbon-abundance-sensitive \texttt{GP}-strong and \texttt{GP}-normal
categories. When calculating the final average iron abundance of
\texttt{GP}-strong stars, no values from $\mbox{[Fe/H]}_{\rm {HP2}}$ were
used. The appearance of a strong G-band at $4300$\,{\AA} may spoil the
\texttt{HP2} measurement. We used $\texttt{GP}>5.0$ as a limit for giants and
$\texttt{GP}>4.0$ for dwarfs. With decreasing iron abundance, it is
increasingly likely that \texttt{GP}-strong stars are very carbon rich (see
also discussion in \S~\ref{carbon_ana}).

Depending on availability of the various $\mbox{[Fe/H]}$ estimates, we
weighted and averaged the values to obtain the final estimate
$\mbox{[Fe/H]}_{final}$. Table~\ref{weights} shows the weights (for dwarfs and
giants seperately) for each of the two iron abundance estimates used to form a
specific, averaged value.  In total we have identified $16\%$ of the sample
stars to be genuinely metal-poor (i.e., 286 objects with
$\mbox{[Fe/H]}<-1.0$). Table~\ref{groups} lists further details on how many
stars were found below a given metallicity. We note that more than half of
these stars are giants (189 out of 286), despite the fact that our sample of
1777 stars comprises mostly dwarfs.

Due to the lower effective temperature, metal-poor giants are better
candidates than dwarfs amongst which to search for e.g., strong r-process
enhancement. The cooler the temperature, the stronger the absorption lines
appear. This is of importance when searching amongst metal-deficient stars
where lines are expected to appear weak. The use of such giants then allows
the tracing of r-process enhancement down to the lowest metallicities.

Taking the number of previously identified metal-poor stars into account, the
numbers of newly-discovered metal-poor objects in this work are shown in
Table~\ref{groups}. For completeness, we also list the number of literature
stars for which a high-resolution analysis has been reported.

We note here that for the most iron-poor star HE~1327$-$2326, a metallicity of
$\mbox{[Fe/H]}_{final}=-4.3$ was obtained, based on its medium-resolution
spectrum. The extremly low metallicity of \mbox{[Fe/H]$=-5.4$} was only
determined from high-resolution spectra. The discrepancy is due to
interstellar Ca blending with the \cak\ line, which could not be resolved in
the medium-resolution data. The overall metallicity covered by our sample
ranges from $\mbox{[Fe/H]}_{final}=-4.3$ up to $\mbox{[Fe/H]}_{final}=-1$
(chosen upper limit).

\subsection{Effective yields and the metallicity distribution
  function}\label{MDF} 

In the present context, the ``effective yield'' describes the fraction of
genuine metal-poor stars below a certain metallicity, compared with the total
number of targeted stars \citep{beers_effyield}. In Table~\ref{eff_yields} we
compare the ``effective yields'' of our bright metal-poor stars with the faint
HES and HK survey stars (data taken from \citealt{ARAA}). We obtain
significantly lower overall effective yields for our work in comparison with
the HES faint sample. This is not unexpected. Saturation effects had an
important (negative) impact on the survey colors by which the bright
metal-poor candidates were selected (see \S~\ref{cal}), and led to the
majority of the entire sample falling outside our criteria for potential
metal-poor stars (see \S~\ref{feh}). Apart from saturation issues, the
majority of the sample is bright and thus likely to be relatively nearby. They
might belong to the thick disk rather than the halo. Hence, the contamination
by more metal-rich stars is larger compared to the sample of the HES faint
stars, which is claimed to comprise $\sim50\%$ stars with $\mbox{[Fe/H]}<-2.0$
\citep{ARAA}. However, when comparing our effective yields with those of the
HK survey (in its initial design with no available colors) they are of the
same order.

\begin{figure}
 \begin{center}
\includegraphics[clip=true]{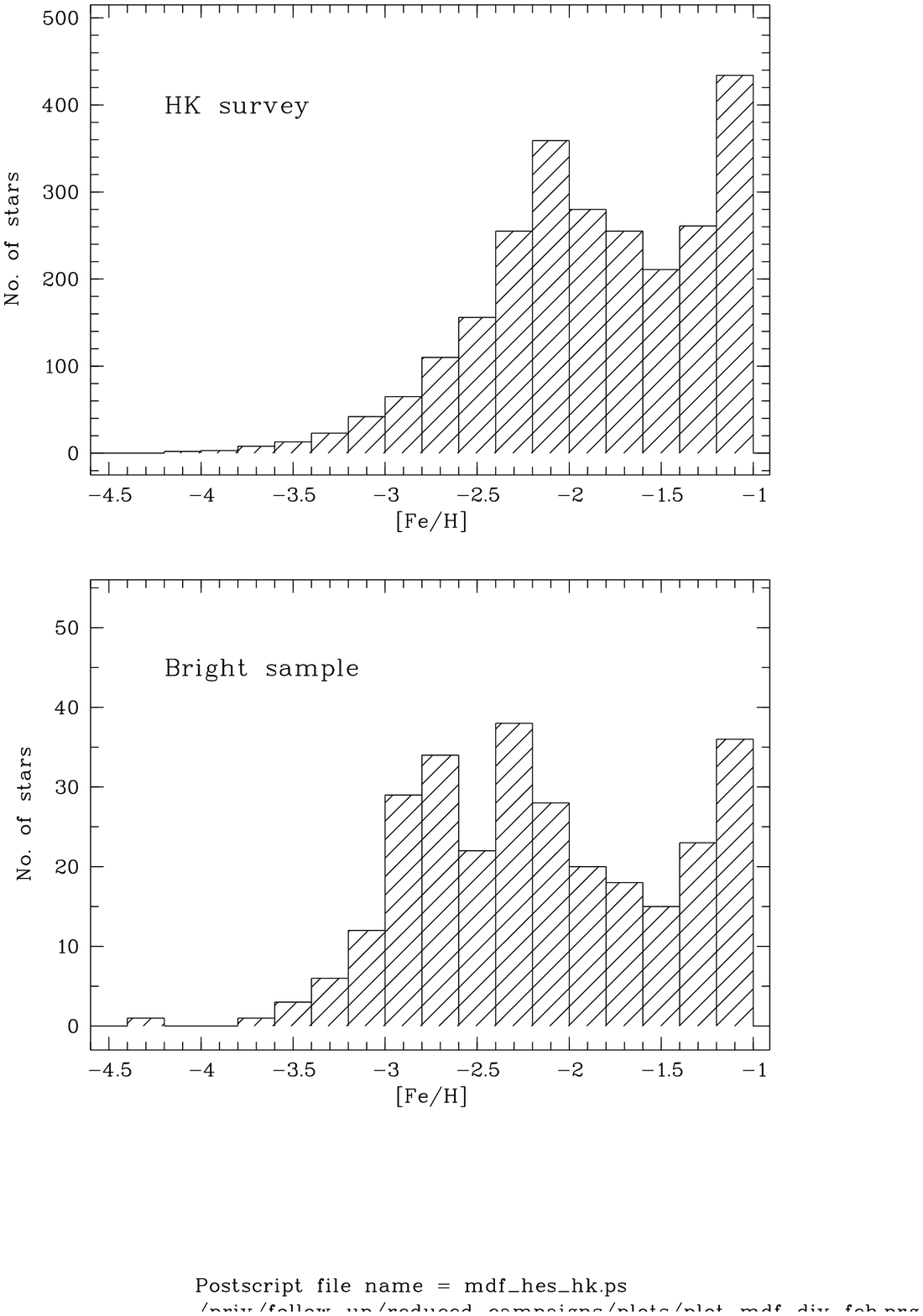}
\figcaption{\label{mdf} Metallicity distribution (MDF) of some 4700 HK survey
  (upper panel) stars compared to our 293 metal-poor stars (lower panel). See
  text for discussion.}
 \end{center}
\end{figure}

The metallicity distribution function (MDF) of the stars with
$\mbox{[Fe/H]}<-1.0$ from the HK survey and our bright star sample is
presented in Figure~\ref{mdf}. Both samples include dwarfs and giants. The HES
bright distribution for stars with $\mbox{[Fe/H]}<-1.0$ has one main peak at
$\mbox{[Fe/H]}\sim-2.3$. This peak likely arises from the fact that the
initial selection of bright stars was aiming to find stars whose \cak\ line
strengths measured in the survey spectra indicated a metallicity of
$\mbox{[Fe/H]}\lesssim-2.5$. Thus, in the case that the selection works well,
the majority of the genuine metal-poor stars confirmed after the
medium-resolution observations should spread around
$\mbox{[Fe/H]}\sim-2.5$. This explanation is supported by the fact that almost
half of the stars with $\mbox{[Fe/H]}<-1.5$ are fainter than $B=13$. Hence,
this subsample does not contain as many of the brighter stars, whose initial
selection as metal-poor candidates was influenced by saturation effects.
However, the dip in the distribution at $\mbox{[Fe/H]}=-2.5$ cannot be
explained in this way. No explanation, other than this may be a low number
statistical effect, can currently be found for this dip.

Towards the higher metallicity cut-off of the MDF the number of stars rises
significantly, likely due to incorrectly-selected objects. There is a link
between a low selection efficiency and the brightness of the objects due to
the saturation effects. About 95\% of the stars in the range of
$-1.5<\mbox{[Fe/H]}<-1.0$ are brighter than $B=13$. This is the brightness
limit at which the saturation becomes much more severe for brighter stars. The
combination of $\mbox{[Fe/H]}$ uncertainties, e.g., from the limits of the
calibrations itself or the loss of sensitivity to metallicity from
$\mbox{[Fe/H]}\sim-1.5$ upwards, likely causes objects with a true
$\mbox{[Fe/H]}>-1.0$ to appear in our MDF as metal-poor stars.

We wish to emphasize that despite the low overall effective yields, this
sample has produced astrophysically important metal-poor stars. Based on our
ongoing high-resolution observations of the most metal-poor stars from this
sample we have found HE~1327$-$2326 ($\mbox{[Fe/H]}<-5.4$) and at least one
strongly r-process enhanced star ($\mbox{[r/Fe]}\sim2.0$, A. Frebel et
al. 2006, in preparation). Results on further stars will be reported elsewhere
(see also \S~\ref{concl}).

\subsection{Carbon abundance estimates}\label{carbon_ana}

Metal-poor stars with enhanced carbon abundance are important to investigate,
for example, the shape of the initial mass function and the formation and
evolution of the first generations of stars. In recent years it has become
apparent that the number of carbon-rich stars increases with decreasing
metallicity (e.g., \citealt{ARAA}). The origin of this trend is not well
understood. Our sample is well suited to investigate the frequency of CEMP
stars amongst metal-poor stars.

\begin{figure*}
 \begin{center}
 \includegraphics[clip=true]{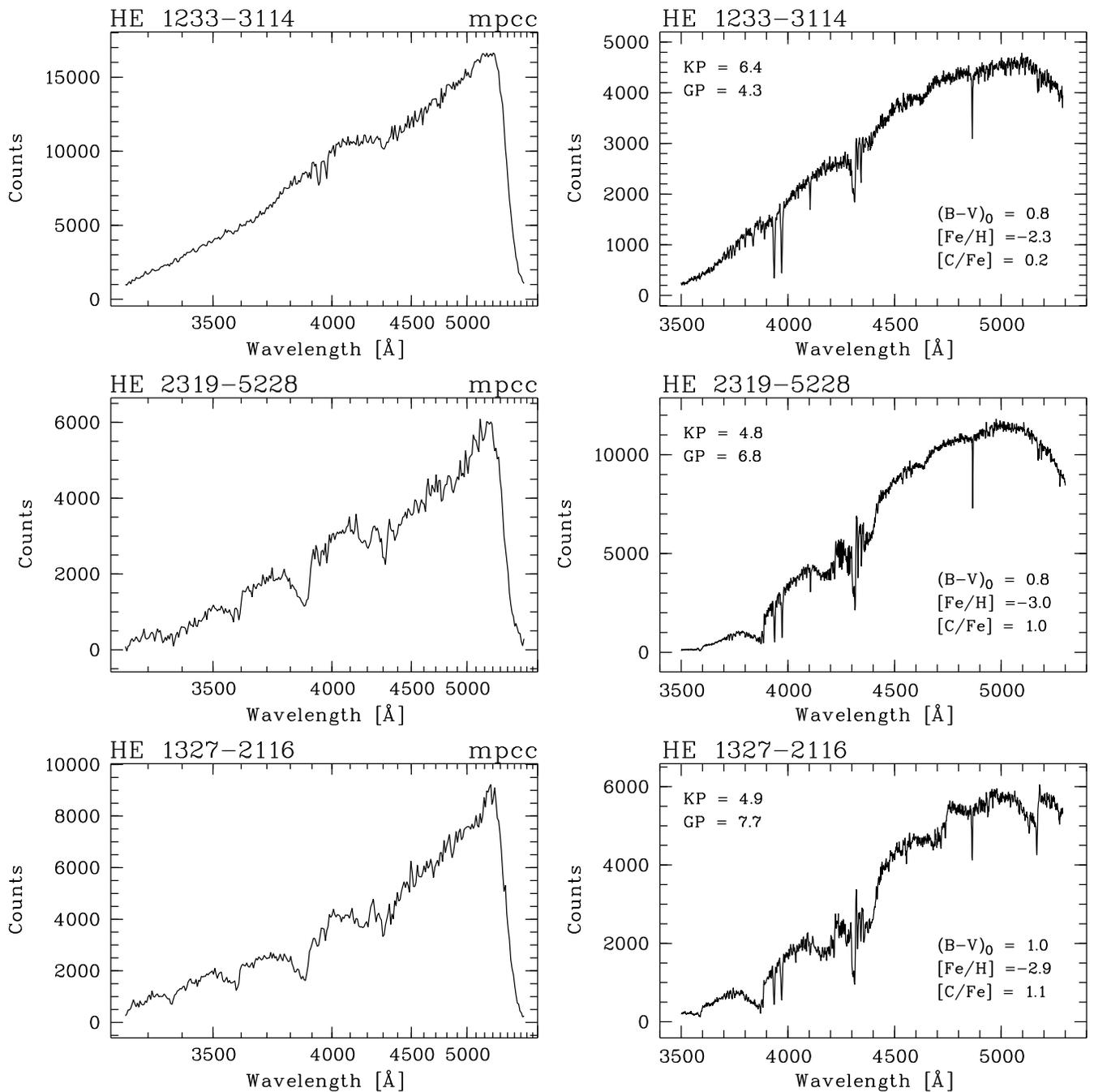}
\figcaption{\label{carbon} Sequence of spectra of \texttt{GP}-strong stars.
 Top panel: The only CH feature visible is the G-band at $\sim4300$\,{\AA}.
 Middle panel: In addition to the G-band, strong CN molecular bands ($3883$
 and $4215$\,{\AA}) are also present. These are already clearly visible in the
 survey spectrum. Bottom panel: The features are even stronger and C$_{2}$
 bands become visible at e.g., $4738$ and $5165$\,{\AA}. }
 \end{center}
\end{figure*}

One way to recognize carbon rich stars among metal-poor objects is by their
G-band strengths (e.g., $\texttt{GP}\gtrsim4.5$). Amongst the 1777 stars in
our sample, $216$ have $\texttt{GP}>5$. Only $30$ stars, however, have
$\texttt{GP}>6$, while $6$ objects have $\texttt{GP}>7$. When considering the
subsample of stars with $\mbox{[Fe/H]}< -1.0$, the numbers become 43, 16, and
6, respectively.  The index begins to saturate from $\texttt{GP}\sim6$ but
certainly, $\texttt{GP}\gtrsim6$ indicates a very strong G-band. If the
metallicity of a star is low and the \texttt{GP} index large, the star is very
likely to be rich in C. Figure~\ref{carbon} illustrates a sequence of
strengths of the G-band and the corresponding C abundance based on the
\citet{rossi_carbon05} calibration (see below).  As an example, we measured a
G-band strength of $\texttt{GP}\sim5.0$ for CS~22892$-$052, the well-studied
r-process- and CEMP star which was rediscovered in our sample. \citet{BPSII}
obtained $\texttt{GP}\sim5.5$ from their medium-resolution spectrum. They also
derive $\mbox{[Fe/H]}= -3.0$, while we obtain $\mbox{[Fe/H]}=
-3.1$. \citet{McWilliametal} then obtained a carbon abundance of
$\mbox{[C/Fe]}\sim 1.0$ (from high-resolution data), which is in good agreement
with our value of $\mbox{[C/Fe]}=1.2$.

\begin{figure}
 \begin{center}
\includegraphics[clip=true]{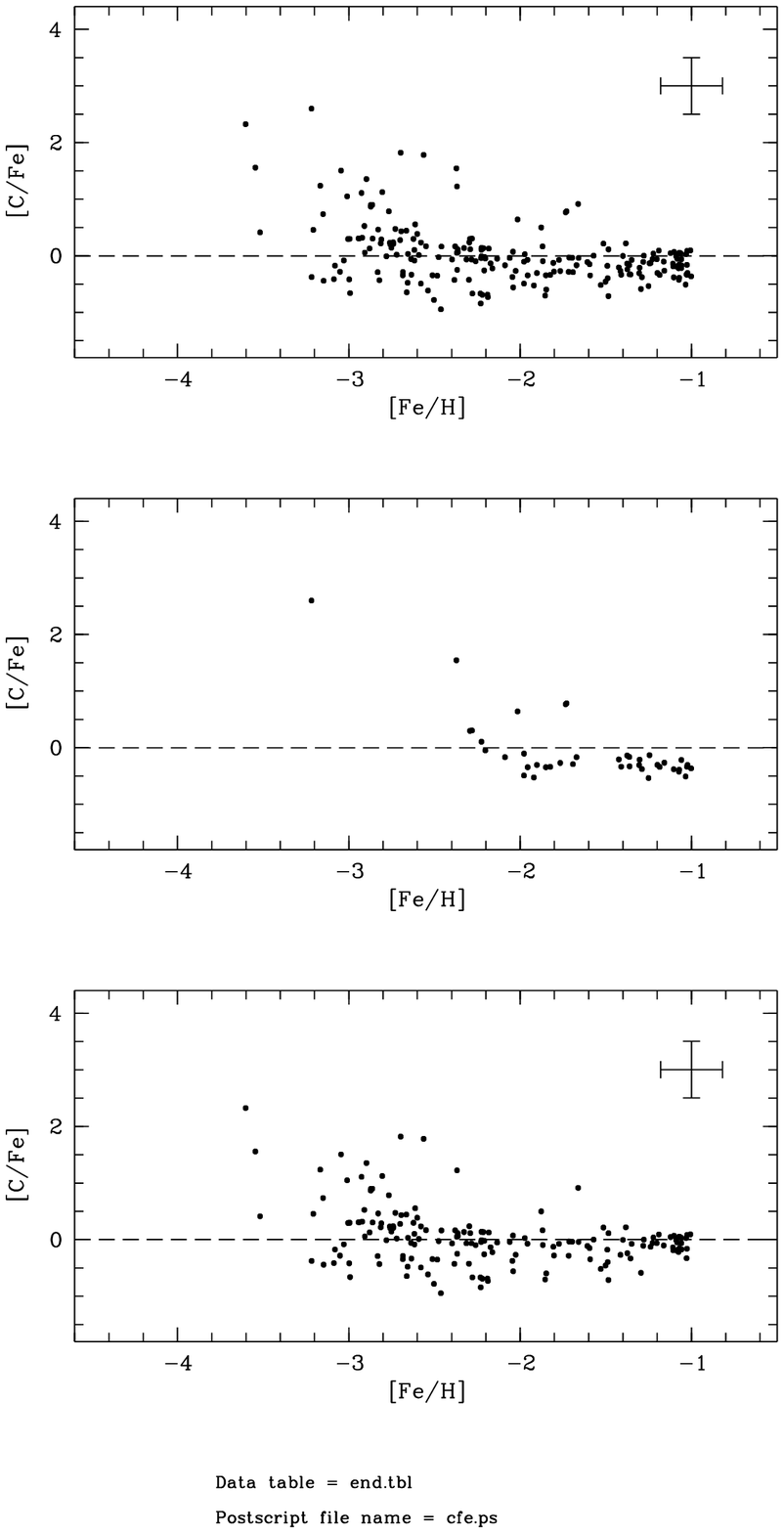}
\figcaption{\label{cfe} $\mbox{[C/Fe]}$ for all stars with $\mbox{[Fe/H]}<-1.0$
  as a function of $\mbox{[Fe/H]}$. An increased spread of $\mbox{[C/Fe]}$
  with decreasing metallicity is clearly seen.}
 \end{center}
\end{figure}

In general, the carbon abundance (with respect to H or Fe) can best be
determined from high-resolution observations. In the absence of such data for
the entire sample of stars with $\mbox{[Fe/H]}< -1.0$ we rely on a different
approach. We use the newly available calibration of \citet{rossi_carbon05} to
obtain estimates for the relative carbon abundances, $\mbox{[C/Fe]}$. Their
regression is suitable for an analysis based on the \texttt{KP} and
\texttt{GP} indices measured in medium-resolution spectra.  We added the
constraint $\texttt{GP}\ge1.0$ to ensure that the G-band was clearly present
in the spectrum.  See Table~\ref{crich} for the numbers of mildly
($\mbox{[C/Fe]}>0.5$) and strongly ($\mbox{[C/Fe]}>1.0$) carbon-enriched
objects found amongst the 53 dwarfs and 121 giants with $\mbox{[Fe/H]}< -2.0$
in our study. The lower limits given in the table are due to the constraint on
the \texttt{GP} index to be larger than 1.0, which introduces a bias against
carbon-rich dwarfs. Figure~\ref{cfe} then shows $\mbox{[C/Fe]}$ for the 234
objects with $\texttt{GP}>1.0$ and $\mbox{[Fe/H]}<-1.0$, plotted against their
iron abundances $\mbox{[Fe/H]}$. An increased spread of carbon abundances with
decreasing iron abundance is clearly seen. However, some stars are missing
from the figure because we have not calculated C abundances for objects with
low G-band strength (i.e., $\texttt{GP}<1.0$). Those stars would likely have
low levels of C enhancement. As a consequence of this constraint on the
\texttt{GP} index most of our metal-poor dwarfs have no carbon abundance
estimate. Due to their higher temperature, the dwarfs are more strongly
affected by this constraint than the giants. Hence, we note here that we did
not attempt to derive a relative frequency of CEMP dwarfs. It is also worth
mentioning that no $\mbox{[C/Fe]}$ was estimated for the most iron poor
dwarf/subgiant HE~1327$-$2326. There is no G-band visible in our
medium-resolution spectrum ($\texttt{GP}=0.5$). Additionally, no stars similar
to HE~1327$-$2326 are available in the calibration \citep{rossi_carbon05}. Any
derived value for such a star would thus be untrustworthy.  Finally, several
authors in the literature have noted a general upper limit of C enhancement
with respect to the Sun only (e.g., \citealt{ryan_ch_envelope},
$\mbox{[C/H]}\sim-1.0$). We also find such an envelope at
$\mbox{[C/H]}\sim-0.7$. However, we note that W. Aoki et al. (2006, in
preparation) and S. Lucatello et al. (2006, in preparation) find an upper
limit on carbon enhancement for carbon-enhanced metal-poor stars close to
[C/H] = 0.0.

\subsection{The frequency of C-enhanced metal-poor stars}

If we now only consider the giants with $\mbox{[Fe/H]}<-2.0$ (which make up
$\sim70\%$ of the objects below that limit), we find the frequency of
metal-poor giants with a strong carbon enhancement of $\mbox{[C/Fe]}\ge1.0$ to
be $\sim9\%\pm2\%$. If we take giants with $\mbox{[Fe/H]}<-2.5$, the frequency
rises to $13\%\pm4\%$, and to $25\%\pm11\%$ amongst giants with
$\mbox{[Fe/H]}<-3.0$. The uncertainties are based on Poisson statistics.  One
potential reason for the low percentage of carbon-enhanced stars below
$\mbox{[Fe/H]} = -2.0$ may be that, for stars with $\texttt{GP}>6.0$, the
calculated $\mbox{[C/Fe]}$ has been underestimated by as much as 0.5 dex
\citep{rossi_carbon05}. We have two stars with $\texttt{GP}=7.5$ and 7.9 which
both have $\mbox{[C/Fe]}=0.9$. Thus, they fall just below our cut-off of
1.0. If those stars were to have an underestimated $\mbox{[C/Fe]}$ by only
0.1\,dex, the frequency of giants with $\mbox{[Fe/H]}<-2.0$ would rise to
$11\%\pm3\%$.

We note here that our frequency estimates are not affected by the constraint
on the \texttt{GP} index (see above). Only stars with $\mbox{[C/Fe]}>1$ were
counted and compared to the total number of stars with $\mbox{[Fe/H]}<-2.0$,
$-2.5$ and $-3.0$, respectively. There clearly is a metallicity effect on the
frequency of CEMP stars. It almost triples when stars with
$\mbox{[Fe/H]}<-3.0$ are compared with those having $\mbox{[Fe/H]}<-2.0$. This
is reflected in the larger spread of $\mbox{[C/Fe]}$ with decreasing iron
abundance.

Our average frequency of CEMP stars of $9\%$, however, is much lower than what
is quoted in \citet{ARAA}. They mention ``at least 20\%'' amongst stars with
$\mbox{[Fe/H]}<-2.0$. For a different sample of HK survey stars with
$\mbox{[Fe/H]}<-2.5$, \citet{marsteller} even find $\sim25\%$ of the stars to
be enhanced in carbon. Our low frequency is, in any case, of the same order as
what was found by \citet{cohen} ($14\%\pm4\%$). S. Lucatello et al. (2006, in
preparation), however, report that at least 20\% of the stars with
$\mbox{[Fe/H]}<-2.0$ from the large HERES sample of \citet{heresII}, based on
high-resolution spectroscopic determinations, exhibit $\mbox{[C/Fe]}>1.0$.

We investigated the apparently low frequency farther by taking the sample of
Beers, Preston \& Shectman (1992; hereafter BPSII) to derive the percentage of
their CEMP stars in the same fashion as adopted in the present work. A
subsample of giants with $\texttt{HP2}<2.0$ was taken and [Fe/H] and [C/Fe]
derived according to the Beers et al. (1999) method and the Rossi et
al. (2005) calibration, respectively. The subset of objects with
$\mbox{[Fe/H]}<-2.0$ and $\mbox{[C/Fe]}>1.0$ was then compared to our sample
after applying the same constraints\footnote{For consistency purposes we
restricted [Fe/H] estimates to the $\mbox{[Fe/H]}_{\rm HP2}$ available for
both samples, avoiding possible systematic differences in the available
colors.}. Despite the different brightness and color distributions, as well as
a different distance distribution (the BPSII sample being a little fainter and
on average further away), we find their sample to have the same, low,
percentage as obtained for ours ($\sim9\%$). We note that by using the Beers
et al. (1999) method to obtain the iron abundance estimates, we are also able
to obtain estimates for the absolute magnitudes and distances of our stars.

\begin{figure}
 \begin{center}
 \includegraphics[width=9cm,clip=true]{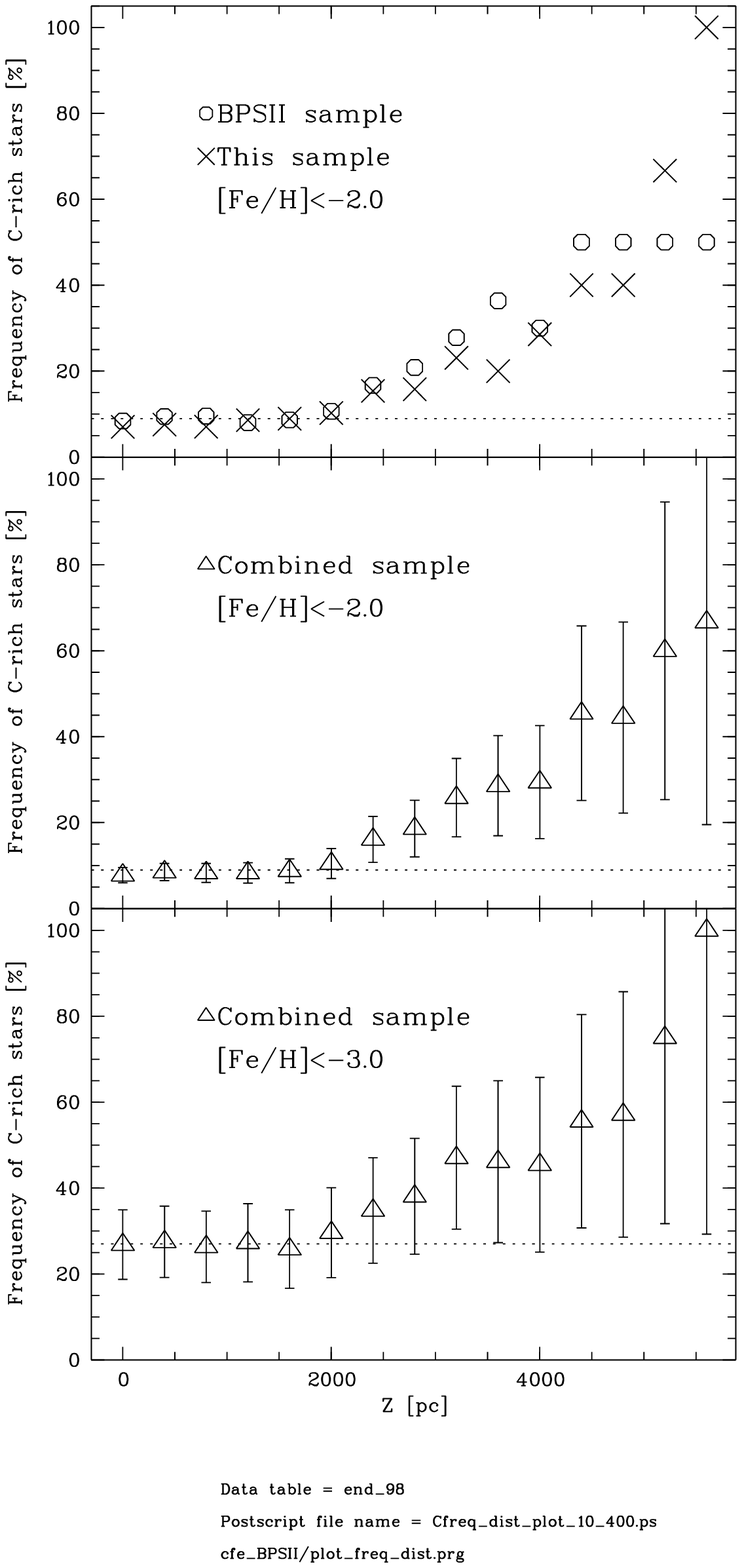}
\figcaption{\label{cfreq_dist} Cumulative frequency of CEMP objects amongst
  metal-poor stars as function of the distance from the Galactic Plane
  $Z$. Note that the cumulative fraction is defined as the number of C-enhanced
  objects amongst metal-poor stars further from the Plane than the indicated
  $Z$. \textit{Top panel}: The frequencies for both our and the BPSII
  sample. \textit{Middle panel}: Combined data set with $\mbox{[Fe/H]}<-2.0$,
  overplotted with Poisson errorbars. \textit{Bottom panel}: Same as middle
  panel, but for subset of stars with $\mbox{[Fe/H]}<-3.0$. See text for
  discussion.}
 \end{center}
\end{figure}

We then tested whether there would be a variation in percentage with distance
$Z$ from the Galactic Plane. Figure~\ref{cfreq_dist} shows the cumulative
fractions of CEMP stars, where the fraction is defined as the number of
C-enhanced objects amongst metal-poor stars further from the Plane than the
indicated $Z$. We performed this for two samples with $\mbox{[Fe/H]}<-2.0$
(top and middle panels) and $\mbox{[Fe/H]}<-3.0$ (bottom panel). The top panel
shows the frequency for both samples seperately. Both data sets exhibit very
similar behavior. The numbers of objects, however, become low with increasing
distance from the Plane. Hence, error bars have only been included in the
middle and bottom panels, where the data sets are combined to obtain the best
estimate for the increase in percentage of CEMP objects for two different
metallicities cut-offs. Concerning the sample with $\mbox{[Fe/H]}<-2.0$, there
is a plateau around $\sim9\%$ out to $Z\sim2$\,kpc. However, as can be seen
from Figure~\ref{cfreq_dist} (middle panel), the frequency begins to increase
significantly when going to larger distances. Considering this behavior, the
apparent discrepancy between our ratio and the Beers and collaborators values
may be resolved if their samples would contain more (fainter) stars at larger
distances from the Plane, compared to our sample. For example, at
$Z\sim3$\,kpc, our figure suggest a frequency of $\sim20-30\%$, in agreement
with their reported estimates.

\begin{figure}
 \begin{center}
\includegraphics[clip=true]{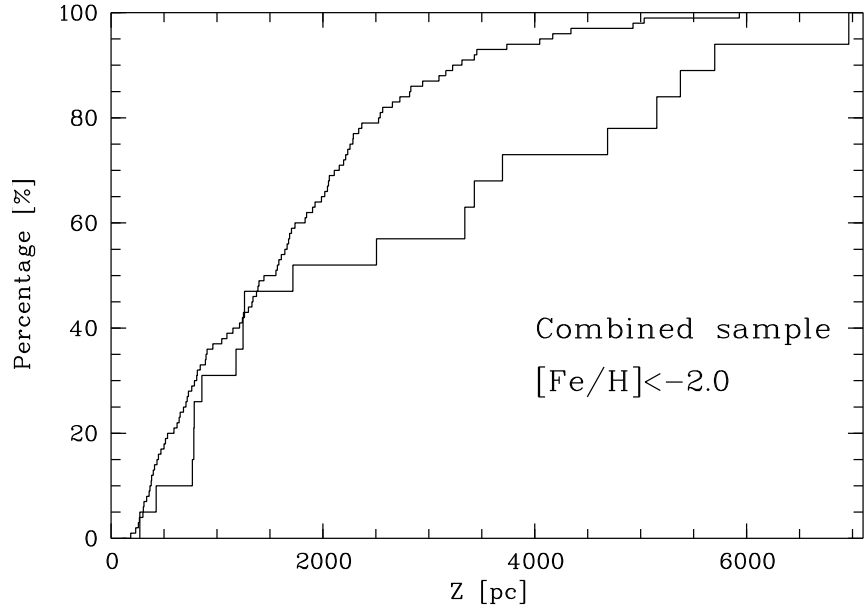}
\figcaption{\label{cfreq_dist_ks} Cumulative percentages of normal metal-poor
    (top curve) and CEMP stars (bottom curve) as a function of distance from
    the Galactic Plane $Z$. Based on the K--S test, the two samples come from
    different populations for $Z\gtrsim2$\,kpc. See text for discussion. }
 \end{center}
\end{figure}

Motivated by the results from Figure~\ref{cfreq_dist}, we investigated the
distribution of our ``normal'' metal-poor stars (i.e., with $\mbox{[Fe/H]}<-2$
and $\mbox{[C/Fe]}<1.0$) with distance from the Plane in order to compare it
with the distribution of the CEMP stars only.  In an attempt to quantify
whether those two samples ('metal-poor' and 'CEMP' samples) were drawn from
the same population, we performed a Kolmogorov-Smirnov (K--S) test
\citep{siegel} on the combined sample as a function of $Z$. The result is
presented in Figure~\ref{cfreq_dist_ks}.

The K--S test rejects the null hypothesis that the two samples are drawn from
the same parent population at the $5\%$ significance level. It is indicative,
however, that for $Z\lesssim2$ the two samples may actually come from the same
population.  Performing additional K--S tests on appropriate subsamples
confirms that up to $Z\sim1.4$\,kpc, it can not be stated that the samples
come from different populations.  For larger $Z$, again, the null hypothesis
is rejected at a $5\%$ singificance level, indicating that the samples are not
drawn from the same population.

From Figures~\ref{cfreq_dist} and \ref{cfreq_dist_ks} it appears that two
effects influence the production of CEMP stars. First, from the middle and
bottom panel of Figure~\ref{cfreq_dist}, we find a relative increase of the
CEMP frequency of $\sim20\%$ when going from the sample with
$\mbox{[Fe/H]}<-2$ to the one with $\mbox{[Fe/H]}<-3$. The bottom panel shows
a plateau around $27\%$, while there is a steep increase with increasing
distance from the Plane. The same pattern is found in the sample with
$\mbox{[Fe/H]}<-2$ (middle panel). Due to low sample numbers, the error bars
are quite large. However, it is indicative that the result would not be
changed significantly if a different sample, containing more lower metallicity
stars, would be employed to improve the statistics. The overall elevated level
for the lower metallicity sample is consistent with previous findings of an
increased frequency of CEMP stars with decreasing [Fe/H]. This behavior may
indicate that it is easier to produce C-rich metal-poor stars in
low-metallicity environments. However, it seems to be independent of the
distance from the Galactic Plane.

The plateau at lower Z seen in Figure ~\ref{cfreq_dist}, together with the
steep increase further out than $Z\sim 2$\,kpc suggests that there is a
second, distance-dependent, effect. This is also indicated by the results from
the K--S tests. At $Z\sim 1.4$\,kpc, a transition occurs between the thick
disk and halo material, causing different chemical signatures (i.e. strong
carbon enrichment) to gain significance in the pool of normal metal-poor
stars.  Estimates of the scale height of the thick disk vary from
$\sim750$\,kpc up to $\sim1.5$\,kpc, where the bulk of the values is between 1
and 1.5\,kpc (e.g., \citealt{norris_thickdisk}, and references therein). The
distance at which we find this transition agrees well with the range of scale
heights available for the thick disk. Our K--S tests supports a transition
with respect to the production of CEMP stars. Beyond this transition distance,
the excess of CEMP objects cannot be directly linked to the formation of
normal metal-poor stars (i.e., the samples do not arise from same
population). At lower $Z$, however, the distance distributions of normal
metal-poor and CEMP stars can not be distiguished, suggesting a common origin
for those two populations as well as for the two chemical elements.
 
The question now arises as to how the productions of the elements (in
particular carbon and iron) are linked to each other and why they change with
distance from the Galactic Plane. Our result indicates that the thick disk
plays a role out to $Z\sim2$\,kpc, but is much less significant beyond
that. Whether or not these differences are due to different mechanisms for the
production of C in the thick disk and the halo remains to be
investigated. Further studies with respect to the (matter) density
distributions of the thick disk and halo populations for different metallicity
ranges would clearly be desired to test our findings more extensively.

\subsection{The catalog of bright metal-poor candidates}

Table~11 presents results for the metal-poor stars identified in this work,
while Table~12 summarizes results for the remaining, non metal-poor stars. The
entire tables are only available electronically.  A portion of each table is
shown for guidance regarding their form and content. 

Both tables share the following columns and are arranged as follows:\\
Columns 1, 2 and 3 lists the HE name, right ascension
and declination (J2000).\\
Columns 4, 5, 6 and 7 list $B$, $B-V$ colors derived from \texttt{HP2}, 
$B-V$ from $(J-K)_{0}$, and $E(B-V)$.\\
Column 8 lists the heliocentric radial velocities.\\
Columns 9, 10 and 11 list the indices \texttt{KP}, \texttt{HP2} and
\texttt{GP}.\\ 

Table~\ref{res_mp} contains additional columns:\\
Column 12 presents the iron abundance $\mbox{[Fe/H]}_{\rm B-V}$ derived
from the \texttt{KP} and the $B-V$ color.\\
Column 13 presents the iron abundance $\mbox{[Fe/H]}_{\rm HP2}$ derived
from the \texttt{KP} and the \texttt{HP2} index.\\
Column 14 presents the iron abundance $\mbox{[Fe/H]}_{\rm Rossi}$ derived from
the 
\citet{rossi_carbon05} calibration.\\
Column 15 presents the iron abundance $\mbox{[Fe/H]}_{final}$, the weighted and
averaged iron abundance of $\mbox{[Fe/H]}_{\rm HP2}$ and
$\mbox{[Fe/H]}_{\rm J-K}$.\\
Column 16 presents the carbon to iron abundance ratio $\mbox{[C/Fe]}$.\\
Column 17 indicates whether the star has been rediscovered as metal-poor
star with $\mbox{[Fe/H]}<-2.0$.

Table~\ref{res_other} also contains Column 12, which shows whether the object
could be identified to be of a certain type of object, such as
horizontal-branch star or emission-line object.

\subsection{Measurement uncertainties}\label{err} 

According to \citet{hespaperIII}, the average dispersion of the HES magnitude
for objects fainter than $B \sim 13$ is $\sigma_{B} \sim 0.2$\,mag. From
Figure~\ref{Bmag}, we determine the dispersion for those fainter stars to be
$\sim0.1$\,mag, with a small offset of $\sim0.1$\,mag. For the brighter
objects, there is an offset of $\sim0.2$\,mag, while the dispersion is
$\sigma_{B} \sim 0.2$\,mag. As already shown in \S~\ref{cal}, the HES $B-V$
color has an uncerainty of 0.1\,mag for stars fainter than $B\sim13$. However,
as is apparent from Figure~\ref{bv_calib}, for brighter stars, the
uncertainties dramatically increase for the brighter stars. Hence, they have
not been used in the determination of the metallicities.  The replacement
$B-V$ color, obtained from the 2MASS $J-K$ color
($\sigma_{J-K}\sim0.03$\,mag), has a typical uncertainty of
$\sigma_{B-V}\sim0.06$\,mag in dwarfs, and $\sigma_{B-V}\sim0.10$\,mag in
giants, when the uncertainties in the regression coefficients are accounted
for.  The average $1\,\sigma$ uncertainty in our radial velocities is
$\sim3$\,km\,s$^{-1}$, based on the two independent measurements.

Using multiple observations ($63$ stars twice, $9$ three times) we tested the
accuracy of our index measurements.  The standard error (based on small number
statistics) of the indices are as follows: $\sigma_{\rm {KP}}= 0.18$,
$\sigma_{\rm {GP}}= 0.15$ and $\sigma_{\rm {HP2}}= 0.20$.
Table~\ref{feh_errors} shows the uncertainties in the different iron abundance
estimates for these changes of the individual input parameters as well as
color changes. We simplified this uncertainty analysis by dividing the sample
into dwarfs ($\texttt{HP2}>2.5$), and giants ($\texttt{HP2}\le2.5$) and
assumed that the uncertainties would not change significantly for members of
each group.  Furthermore, we divided each group into the more metal-poor stars
($\mbox{[Fe/H]}<-2$) and more metal-rich ones ($\mbox{[Fe/H]}>-2$), to test
whether there would be an effect with metallicity. Most uncertainties do not
significantly differ between dwarfs and giants for each metallicity group,
execpt in two cases (see Table~\ref{feh_errors} for details.)

Generally, the uncertainties are small, around \mbox{$\sim0.1$}\,dex per
parameter change. It is apparent, though, that the major source of error
arises from uncertainty in the $B-V$ color. For the more metal-poor objects it
is $\sim0.2$\,dex, and doubles for stars with $\mbox{[Fe/H]}>-2$. This,
however, is in agreement with the calibration losing sensitivity above
$\mbox{[Fe/H]}\sim-1.7$ \citep{BeersCaKII}. To obtain reliable iron abundance
estimates, other techniques, such as the ACF, would have to be employed in
this metallicity range (see Beers et al. 1999). In any case, this behaviour
argues for the use of the \texttt{HP2} derived abundances. Even in the range
where this method begins to lose sensitivity, (i.e., for cooler giants), the
uncertainty for giants is still lower than that for the $B-V$ color. For stars
with $\texttt{KP}<2$\,{\AA}, the uncertainty in [Fe/H] increases by a factor
of $\sim2-3$ compared to the typical value.

We are concerned that our uncertainties may be underestimated. This is
apparent from our uncertainties for the [Fe/H] and [C/Fe] estimates based on
the Rossi et al. (2005) calibration. Our derived values appear to be rather
small, but when the uncertainties of the regression coefficients are taken
into account, they increase to $0.5$ ($\mbox{[Fe/H]}<-2$) and $0.6$\,dex
($\mbox{[Fe/H]}>-2$), where the dwarfs have $\sim0.15$\,dex lower values than
the giants. A similar, but less strong effect occurs for [C/Fe], where the
uncertainties raise to $\sim0.2$\,dex for all stars.

\begin{figure}
 \begin{center}
 \includegraphics[clip=true]{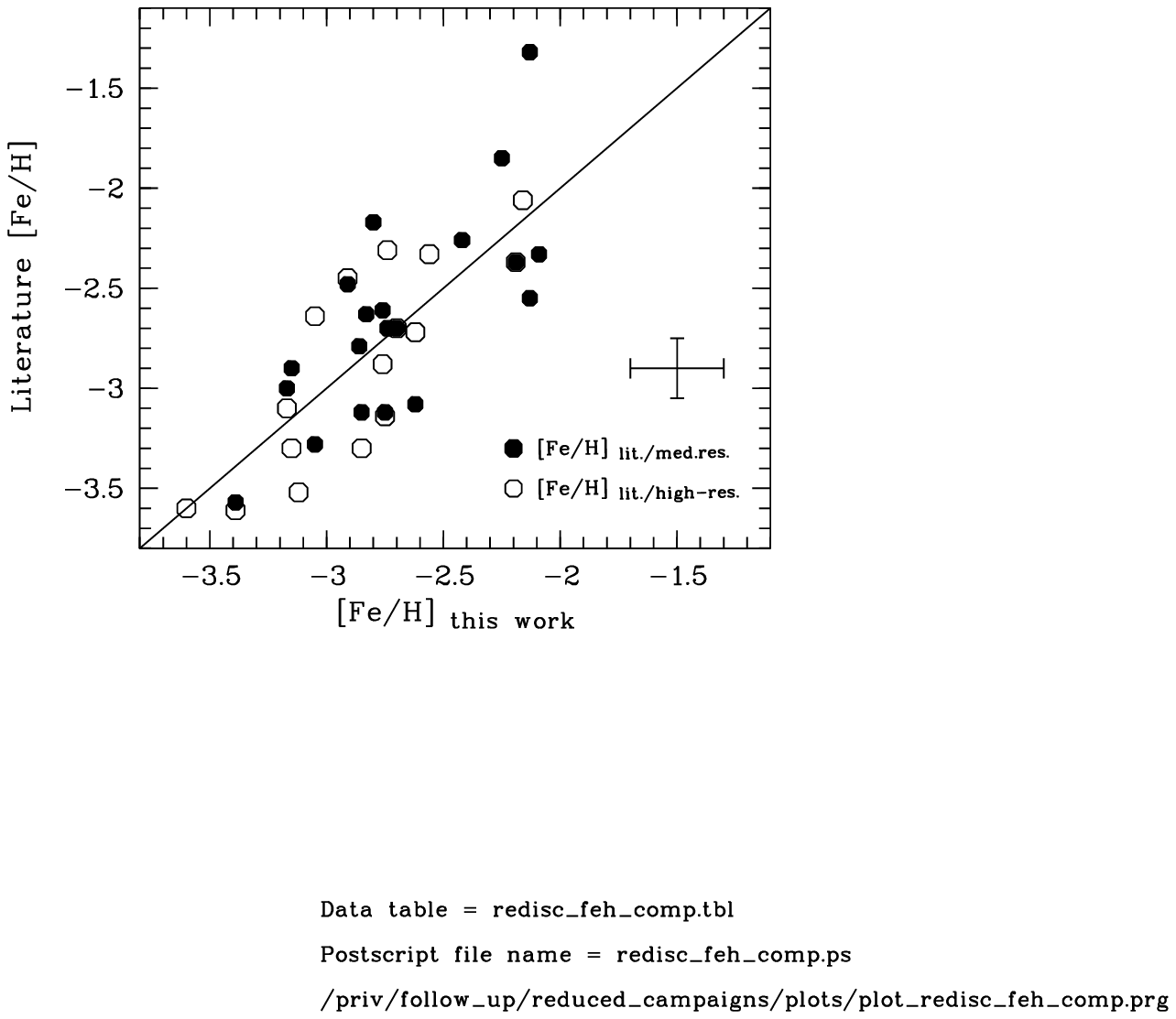}
\figcaption{\label{redisc_plot} Comparison of our final metallicity estimates
  for rediscovered stars with values derived from medium- (\textit{filled
  circles}) and high-resolution (\textit{open circles}) spectra available in
  the literature (see Table~\ref{rediscoveries} and text for further
  details).}
 \end{center}
\end{figure}

In Figure~\ref{redisc_plot} a comparison is shown between our iron abundance
estimates and literature values for the set of rediscovered stars. The mean
deviation from the one-to-one relation is $\sim0.2$\,dex when comparing with
the high-resolution literature data and $\sim0.3$\,dex for the
medium-resolution literature values. These are in good agreement with the
$\mbox{[Fe/H]}$ uncertainties discussed above.

\subsection{Other findings} \label{gen_fin}

Although we were searching for metal-poor stars, we coincidentally found
several groups of other objects. The majority of these other objects are too
hot or of some other nature to be identified as metal-poor in the present
survey. Due to the sometimes very prominent \ion{He}{1} lines in their
spectra, we were able to visually identify 17 O- and B-stars. These objects
are likely to have been selected due to their apparently weak \cak\ line
strength resulting from their high temperatures, in combination with erroneous
HES color information (see also the caption of Table~\ref{res_other}). Also,
about 330 stars have $\texttt{HP2}$ indices larger than 5.3, which is our
upper limit for metal-poor main-sequence turnoff stars. A further 360 stars
have $4.0<\texttt{HP2}<5.3$, for which we did not compute iron abundances
(they either did not satisfy all the criteria for the $\mbox{[Fe/H]}$
determination, or they were simply too metal-rich). $\texttt{HP2}=4.0$
corresponds to $B-V\sim0.4$. We regard these objects as either main-sequence
A-type or field horizontal-branch candidates. At least two stars in our
sample for which we initially computed metallicities have been identified as
A-type stars (CS~22185-008, CS~22185-025; \citealt{wilhelm99,
beers96_fhb}). They both have $\texttt{HP2}>5.3$.
This suggests that many A-type stars may exist among these almost 700
objects.

\begin{figure}
 \begin{center}
 \includegraphics[clip=true]{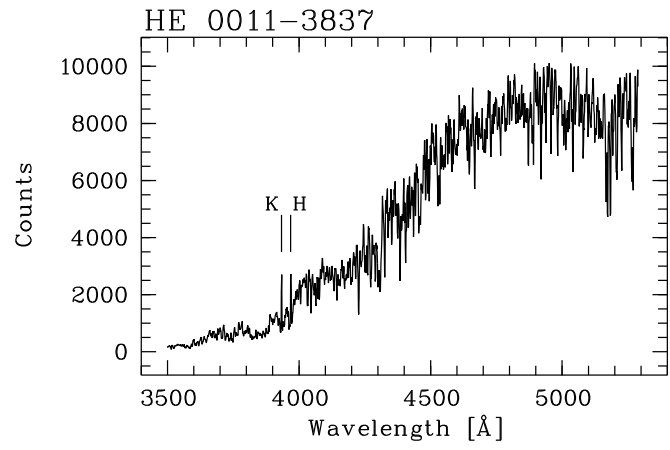}
\figcaption{\label{emission} Example of an emission-line object found in our
  sample. As CS~31077-034, this star was classified by \citet{beers_em2} to
  have ``moderate'' \ion{Ca}{2} H and K emission.}
 \end{center}
\end{figure}

Many of the cooler stars found in the sample exhibit strong \ion{Ca}{1}
absorption at $4226$\,{\AA} in their spectra. Seventeen stars exhibit emission
cores in their \ion{Ca}{2} H and K lines, some of which also show emission in
the H$\delta$ line. From our cross-correlation with the HK survey, we know of
two such stars previously identified as emission-line
objects. \citet{beers_emission, beers_em2} list emission-line HK objects which
were found in the same fashion as ours, as by-products in the search for
metal-poor stars.  HE~1521$-$0033 (CS~22890-077= BS~16559-084;
\citealt{beers_emission}) was marked to have ``weak \ion{Ca}{2} H and K core
emission'' while HE~0011$-$3837 (CS~31077-034; \citealt{beers_em2}) was
classified as having ``moderate \ion{Ca}{2} H and K emission''.  Emission-line
cores result from active chromospheres in late-type stars, which may be used to
trace the structure of the outer stellar atmosphere. Figure~\ref{emission}
shows the spectrum of one such star.  Further details on these objects can be
found in Table~\ref{res_other}.

It is beyond the scope of this investigation to track how many of our
emission-line objects and field horizontal-branch stars have been previously
identified by other studies. We believe, though, that there would be a number
of stars that are already known in the literature.

Through our search with the SIMBAD database we also found that our sample
contains at least three variables (of which two are RR Lyrae objects), one
eclipsing binary, and one nova-like star. Another object is a double or
multiple star. From selected high-resolution observations we know of at least
two double-lined spectroscopic binaries. We marked those stars accordingly in
Table~\ref{res_other} (column (12)).

\section{Summary and outlook on future high-resolution
  spectroscopy}\label{concl}  

In this paper we have presented a sample of 1777 bright ($9<B<14$) metal-poor
candidates selected from the Hamburg/ESO survey. Compared to the previously
explored faint HES metal-poor stars, our sample contains stars whose
objective-prism spectra suffer from saturation effects. The selection
procedure has been described in detail, together with results from
medium-resolution follow-up spectroscopy of the entire sample
(Table~\ref{res_mp}). From the medium-resolution observations, the metallicity
$\mbox{[Fe/H]}$ was determined by means of the \citet{BeersCaKII} method
involving the \texttt{KP} index and a spectroscopically or photometrically
derived $B-V$ color. Based on the 2MASS $J-K$ colors, we also employed the
\citet{rossi_carbon05} calibrations to obtain a third iron abundance estimate,
as well as [C/Fe]. Where available, the iron abundances derived from the
different methods were weighted and averaged. In total we have identified
$16\%$ of the sample stars to be of metal-poor nature (i.e., 286 objects with
$\mbox{[Fe/H]}<-1.0$). Of those, $174$ have $\mbox{[Fe/H]}<-2.0$, $98$ have
$\mbox{[Fe/H]}<-2.5$, and $23$ have $\mbox{[Fe/H]}<-3.0$. Amongst those
objects with $\mbox{[Fe/H]}<-2.0$, we found 13 carbon-rich objects with
$\mbox{[C/Fe]}>1.0$ . Apart from the metal-poor stars, a few examples of
other groups were also serendipitously identified (through the SIMBAD data
base), such as main-sequence A-type stars, emission-line objects, variable
stars and one nova-like star.

Due to rediscoveries of several metal-poor stars from the HK survey and other
studies, $\sim30$ of the 286 metal-poor stars are already known as such in the
literature. This leaves $\sim250$ bright objects newly classified as
metal-poor. The effective yields of our sample are of roughly the same order
as found in the HK survey. These are significantly lower than reported for the
faint HES stars. This difference can largely be understood in terms of the
saturation effects and the subsequent, adversely affected, selection. We also
investigated the frequency of C-enhanced objects amongst metal-poor stars with
distance from the Galactic Plane. Two effects are likely to play a role in the
production of C-rich objects in the Galaxy. First, the frequency of CEMP stars
is higher for lower-metallicity samples, perhaps indicating that the carbon
production becomes more efficient with decreasing [Fe/H]. We find this effect
to exist independent of distance from the galactic plane. The second effect is
a plateau of the frequency of CEMP objects out to $Z\sim2$\,kpc (regardless of
the sample metallicity cut-off), complemented with a steep increase for larger
distances. This may be due to a different mechanism for the production of
carbon and iron in the different populations. For the halo population, at
least, the mechanism does not seem to be the same for normal metal-poor and
CEMP stars.

Using 4-8m class telescopes snapshot spectroscopy ($R \sim 20,000$, $S/N > 30$
per pixel at $\sim4100$\,{\AA}) of selected targets with sufficiently low
metallicity is well underway. Our general procedure is as follows. If a
spectrum indicates any interesting features, e.g., r-/s-process enrichment or
a metallicity as low as $\mbox{[Fe/H]}<-4.0$ we are obtaining higher
resolution (R$>40,000$), higher signal-to-noise observations. These data are
then used for a detailed abundance analysis. This has already been done for
one bright star in particular, HE~1327$-$2326. Subaru/HDS data were obtained
and revealed it to be the new record holder for the most iron-poor star known
to date, with $\mbox{[Fe/H]}=-5.4$ (Frebel et al. 2005, 2006;
\citealt{Aokihe1327}). Further papers of this series on bright metal-poor
stars from the HES will include the high-resolution spectroscopic analyses of
the metal-poor objects indentified in the present study as well as the
candidate follow-up spectroscopy of the remaining, yet unexplored, 51 HES
fields (Frebel et al. 2006, in preparation).

\acknowledgments We thank S. Tsangarides and J.~D. Tanner for reducing the AAT
data and E. Westra for useful comments on earlier versions of the
manuscript. A.~F., J.~E.~N. and M.~S.~B. acknowledge support from the
Australian Research Council under grants DP0342613 and DP0663562. N.~C. and
D.~R. acknowledge support from Deutsche Forschungsgemeinschaft under grants
Ch~214/3 and Re~353/44. T.~C.~B acknowledges support from grants AST 04-06784
and PHY 02-16783, Physics Frontier Centers/JINA: Joint Institute for Nuclear
Astrophysics, awarded by the U.S. National Science Foundation. S.~R. thanks
FAPESP, CNPq and Capes for partial financial support.

This research has made use of the SIMBAD database, operated at CDS,
Strasbourg, France. It makes use of data products from the Two Micron All Sky
Survey, which is a joint project of the University of Massachusetts and the
Infrared Processing and Analysis Center/California Institute of Technology,
funded by the National Aeronautics and Space Administration and the National
Science Foundation.
 
Facilities: \facility{ATT (DBS)}, \facility{CTIO:1.5m (R-CS)}, \facility{Blanco
(R-CS)}, \facility{Mayall (R-CS)}, \facility{ESO:3.6m (EFOSC)}, \facility{AAT
(UCLES)}


\clearpage


\begin{deluxetable}{rrl}
\tablecaption{\label{selection_results}Results of the visual inspection}
\tablewidth{0pt}
\tablehead{
\colhead{Class$^{a}$}& \colhead{Number} & \colhead{Comment}}
\startdata
   ``\texttt{mpca}''&     9 & No \cak\ visible \\
   ``\texttt{unid}''&    84 & Uncertain if \cak\ visible\\
   ``\texttt{mpcb}''&   248 & Relatively weak \cak\ \\
   ``\texttt{mpcc}''&  1426 & Relatively strong \cak\
\enddata
\tablecomments{1767 objects were selected from the $5081$ raw candidates.}
\tablenotetext{a}{See Figure~\ref{mpc} for example objects which illustrate the
  different classes.}
\end{deluxetable}


\begin{deluxetable}{rr}
\tablecaption{\label{num_id}Bright candidates identified in other catalogs} 
\tablewidth{0pt}
\tablehead{
\colhead{Catalog}& \colhead{Number} }
\startdata 
HK survey; \citet{BeersCaKII}& 26 \\
\citet{Ryanetal:1991} & 9    \\
\citet{carney_cat}    & 5    \\
\citet{Bidelman_MacConnell}& 1 \\
SIMBAD (i.e., \textit{any} other cat.) & $\sim1250$\tablenotemark{a}\\
2MASS; \citet{2MASS}  & 1568 \\
SPM3.1; \citet{SPM}   & 392  \\
UCAC2; \citet{UCAC}   &1717 
\enddata 
\tablenotetext{a}{Of these stars, actually only $\sim$150 have at least one
  reference listed in SIMBAD.} 
\end{deluxetable}

\begin{deluxetable}{lllllll}
\tabletypesize{\scriptsize}
\tablecaption{\label{rediscoveries} Previous identifications of stars with
 $\mbox{[Fe/H]}_{\rm final}<-2.0$}    
\tablewidth{0pt} 
\tablehead{ 
\colhead{HE name}& 
\colhead{$\mbox{[Fe/H]}_{\rm final}$}& 
\colhead{Other names\tablenotemark{a}}&
\colhead{$\mbox{[Fe/H]}_{\rm {med.-res.}}$}&
\colhead{$\mbox{[Fe/H]}_{\rm {high-res.}}$}& 
\colhead{References}}
\startdata
HE~0002$-$3233 & $-$2.86 & CS~22961-023                       &$-$2.79& \nodata& 1  \\
HE~0007$-$1752 & $-$2.16 & CS~31060-043                       &\nodata& $-$2.06& 2, 3  \\
HE~0017$-$3646 & $-$2.42 & CS~30339-040                       &$-$2.26& \nodata& 4   \\
HE~0027$-$1221 & $-$2.56 & CS~31062-050                       &\nodata& $-$2.33& 2, 5  \\
HE~0039$-$2635 & $-$3.60 & CS~29497-034                       &\nodata& $-$3.60& 6    \\
HE~0054$-$2542 & $-$3.05 & CS~22942-019                       &$-$3.28& $-$2.64& 4, 5  \\
HE~0056$-$3022 & $-$3.15 & CD$-$30 298, a, b, c, d, e       &$-$2.90& $-$3.30& 7, 8  \\
HE~0111$-$1118 & $-$2.13 & CS~22174-007                       &$-$2.55& \nodata& 2, 1\\ 
HE~0236$-$0242 & $-$2.13 & CS~22954-015                       &$-$1.32& \nodata& 4   \\    
HE~0311$-$1046 & $-$3.39 & CS~22172-002                       &$-$3.57& $-$3.61& 9, 10  \\
HE~0315$-$1528 & $-$2.91 & CS~22185-007                       &$-$2.48& $-$2.45& 4, 3  \\
HE~0317$-$3301 & $-$2.85 & CD$-$33 1173, a, b, c, d, e, g   &$-$3.12& $-$3.30& 11, 12  \\
HE~0336$-$2412 & $-$2.74 & CD$-$24 1782, b, c, d, e         &$-$2.70& $-$2.31& 7, 13  \\
HE~0409$-$1212f& $-$2.62 & CS~22169-035                       &$-$3.08& $-$2.72& 4, 14 \\
HE~0926$-$0508 & $-$2.80 & BS~17572-100                      &$-$2.17& \nodata& 1  \\
HE~0938+0114   & $-$2.70 & BD+01 2341p, G 48-29, LP 608- 62 &$-$2.70& $-$2.70& 15, 16  \\
HE~1144$-$1349 & $-$2.75 & BD$-$13 3442, a, b, c, g         &$-$3.12& $-$3.14& 11, 17  \\
HE~1208+0040   & $-$2.19 & G 11-44, Ross 453, c, d, g       &$-$2.37& $-$2.37& 11, 18  \\
HE~1337+0012   & $-$3.12 & G 64-12                          &\nodata& $-$3.52& 19 \\
HE~2002$-$5843 & $-$2.76 & CS~22873-128                      &$-$2.61& $-$2.88& 4, 20 \\
HE~2119$-$4653 & $-$2.83 & [M93]16090                       &$-$2.63& \nodata& 21  \\
HE~2138$-$1526 & $-$2.25 & CS~22944-11                       &$-$1.85& \nodata& 4   \\
HE~2214$-$1654 & $-$3.17 & CS~22892-52                       &$-$3.00& $-$3.10& 4, 20 \\ 
HE~2301$-$0248 & $-$2.09 & G28-31, LP 701-67, b, h          &$-$2.33& \nodata& 22  
\enddata

\tablecomments{$\mbox{[Fe/H]}_{\rm {med.-res.}}$ and $\mbox{[Fe/H]}_{\rm
    {high-res.}}$ refer to the iron abundances based on medium- and high-resolution
spectra first found in the literature. The corresponding references are given.}
\tablenotetext{a}{Selected other names only; stars are also found in the
  following catalogs: a: PPM \citep{ppm}, b: GSC 
\citep{gsc}, c: Tycho/Hipparcos \citep{hip_tycho}, d: CPD \citep{cpd}, e:
Hipparcos Input Cat. \citep{hic}, f: SAO, h: NLTT \citep{luytennltt} }
\tablenotetext{\ }{References. ---
(1) \citet{schuster04},
(2) \citet{norris85}, 
(3) \citet{lai}, 
(4) \citet{BPSII}, 
(5) \citet{aoki_lead2002}, 
(6) \citet{barbuy}, 
(7) \citet{Bond1980},
(8) \citet{bessell_sutherland},
(9) \citet{ryan96}, 
(10) \citet{Norrisetal:2001},
(11) \citet{Ryanetal:1991}, 
(12) \citet{spite_spite93},
(13) \citet{luck_bond83}, 
(14) \citet{honda04}, 
(15) \citet{hobbs87},
(16) \citet{pilachowski93},
(17) \citet{ryanSDIV91},
(18) \citet{thorbrun94},
(19) \citet{1981G64-12}, 
(20) \citet{McWilliametal}, 
(21) \citet{morrison93}, 
(22) \citet{lairdpm88}}
\end{deluxetable}

\begin{deluxetable}{llllll}
\tablecaption{\label{other_redisc} Selected previous identification of other objects}    
\tablewidth{0pt} 
\tablehead{ \colhead{HE name}& 
\colhead{Other names\tablenotemark{a}}& 
\colhead{Ref.} & \colhead{Comment}} 
\startdata
HE~0411$-$3558 & CS~22186-005               & 1& HB \\
HE~2247$-$4113 & CD-41 15048, a, b, c, d, e& 2& RHB \\
HE~1252$-$0511 & AT Vir, b, c, e        & 3& RR Lyr Var \\
HE~2325$-$4743 & RV Phe, CD-48 14514, c & 4& RR Lyr Var\\
HE~2210$-$4058 & SON 7481              & 5& suspected RR Lyr Var
\enddata
\tablecomments{Selected other names; selected references which first referred
  to the nature of the star.}
\tablenotetext{a}{Also found in the following catalogs: a: PPM \citep{ppm}, b:
  GSC \citep{gsc}, c: Tycho/Hipparcos \citep{hip_tycho}, d: CPD \citep{cpd}, e:
Hipparcos Input Cat. \citep{hic}}
\tablenotetext{\ }{References. ---
(1) \citet{BPSII}, 
(2) \citet{norris85}, 
(3) \citet{hoffmeister},
(4) \citet{lourens}, 
(5) \citet{demartino}}
\end{deluxetable}

\begin{deluxetable}{lllllll} 
\tablecaption{\label{non_mp_resdisc} Stars with
 $\mbox{[Fe/H]}<-2.5$ not identified as metal-poor in the visual inspection} 
\tablewidth{0pt} 
\tablehead{ 
\colhead{HE name}& 
\colhead{Classification}&
\colhead{Other names}&
\colhead{$\mbox{[Fe/H]}$}& 
\colhead{Type}&
\colhead{Ref.}}
\startdata
 HE~0038$-$2423&\texttt{hbab}&CS~29497-030        & $-$2.8 &  TO (hot) & 1 \\
 HE~0042$-$2141&\texttt{hbab}&CS~29527-057        & $-$2.7 &  HB       & 2 \\
 HE~0303$-$2230&\texttt{star}&LP 831-70           & $-$3.5 &           & 3 \\ 
 HE~2134$-$3940&\texttt{ovl} &CS~22948-027        & $-$3.2 & C-rich TO & 4 \\
 HE~2131+0010  &\texttt{unid}&G26-12, LP 638-7 & $-$2.6 & TO & 4 
\enddata
\tablecomments{For details on the classifications, see text. 
TO: Turnoff star,
HB: Horizontal Branch Star}
\tablenotetext{\ }{References. ---
(1) \citet{sivarani},
(2) \citet{schuster04},
(3) \citet{Ryanetal:1991},
(4) \citet{barbuy}}
\end{deluxetable}

\begin{deluxetable}{lllr}
\tablecaption{\label{med-res}Telescopes used for the medium resolution observations}
\tablewidth{0pt}
\tablehead{
\colhead{Telescope}& \colhead{Instrument} & \colhead{Observers}&
\colhead{No.}}
\startdata
2.3\,m, SSO    & DBS              & Bessell, Frebel, & 1477\\
               &                  & \multicolumn{1}{l}{Norris, Thom} &\\
4.0$\,$m, KPNO & R-C Spectrograph & Beers, Marsteller & 92\\ 
1.5$\,$m, CTIO & R-C Spectrograph & Rhee          & 82 \\
3.6$\,$m, ESO  & EFOSC2           & Fechner       & 67 \\
4.0$\,$m, CTIO & R-C Spectrograph & Beers, Rossi  & 46 \\
3.9$\,$m, AAO  & RGO              & Norris        & 13 
\enddata
\end{deluxetable}

\begin{deluxetable}{llll}
\tablecaption{\label{weights} Weights applied to the iron abundances}
\tablewidth{0pt}
\tablehead{
\colhead{Iron abundances} &\colhead{} &\colhead{Weights} &\colhead{}}
\startdata
$\mbox{[Fe/H]}_{\rm {HP2}}$, $\mbox{[Fe/H]}_{\rm {B-V}}$   & dwarfs & 0.8, 0.2 &\\
                                                               & giants & 0.5, 0.5 &\\
$\mbox{[Fe/H]}_{\rm {B-V}}$, $\mbox{[Fe/H]}_{\rm {Rossi}} $& dwarfs & 0.5, 0.5 &\\ 
                                                               & giants & 0.5, 0.5 &\\
$\mbox{[Fe/H]}_{\rm {HP2}}$, $\mbox{[Fe/H]}_{\rm {J-K}}$& dwarfs & 0.7, 0.3 &\\
                                                            & giants & 0.5, 0.5 &
\enddata
\tablecomments{``dwarfs'' refers to $2.5<\texttt{HP2}<5.3$ and ``giants'' to
  $0.25<\texttt{HP2}\le2.5$. See text for further details.}
\end{deluxetable}

\begin{deluxetable}{lrrrr}
\tablecaption{\label{groups}Metal-poor stars found in the sample}
\tablewidth{0pt}
\tablehead{
\colhead{$\mbox{[Fe/H]}$ range} &\colhead{$N$}&
\colhead{$N_{\rm {redisc.}}$\tablenotemark{a}} & 
\colhead{$N_{\rm {high-res.}}$\tablenotemark{b}} &
\colhead{$N_{\rm {new}}$} } 
\startdata
$\mbox{[Fe/H]}<-2.0$   & 174 &  29 &  16 & 145 \\
$\mbox{[Fe/H]}<-2.5$   &  98 &  19 &  14 &  79 \\
$\mbox{[Fe/H]}<-3.0$   &  23 &   6 &   6 &  17
\enddata
\tablenotetext{a}{The rediscoveries refer to previous identifications of stars
  with $\mbox{[Fe/H]}<-2.0 $ listed in SIMBAD.}
\tablenotetext{b}{These stars are rediscovered and have high-resolution
  spectral analysis reported in the literature. See Table~\ref{rediscoveries}
  for references and text for discussion.}
\end{deluxetable}

\begin{deluxetable}{lrrrr}
\tablecaption{\label{eff_yields} ``Effective yields'' of the sample}  
\tablewidth{0pt} 
\tablehead{ \colhead{Survey}& \colhead{No.} &
\multicolumn{3}{c}{$\mbox{[Fe/H]}$}\\ \colhead{}& \colhead{}&
\colhead{$<-2.0$}& \colhead{$<-2.5$}& \colhead{$<-3.0$}} \startdata
HK survey/no $B-V $       & 2614 & 11\% & 4\% &1\% \rule{0ex}{2.3ex}\\
HK survey/with $B-V $     & 2140 & 32\% &11\% &3\%  \rule{0ex}{2.3ex}\\
\textbf{HES bright (all)}&\textbf{1777}&\textbf{11\%}&\textbf{6\%}&\textbf{2\%} \\
\textbf{HES bright ($B\ge13$)}&\textbf{161}&\textbf{49\%}&\textbf{31\%}&\textbf{11\%} \\
\textbf{HES bright ($B\ge13$) dwarfs}&\textbf{155}&\textbf{33\%}&\textbf{16\%}&\textbf{2\%} \\
\textbf{HES bright ($B\ge13$) giants}&\textbf{106}&\textbf{58\%}&\textbf{39\%}&\textbf{15\%} \\
HES faint (dwarfs)        & 571  & 59\% &21\% &6\%  \rule{0ex}{2.3ex} \\
HES faint (giants)        & 643  & 50\% &20\% &6\%  \rule{0ex}{2.3ex}
\enddata
\tablecomments{Results for HK survey and faint HES giants are taken from
  \citet{ARAA}.}
\end{deluxetable}

\begin{deluxetable}{llr}
\tablecaption{\label{crich}CEMP stars with $\mbox{[Fe/H]}<-2.0$ found in the
  sample} 
\tablewidth{0pt}
\tablehead{\multicolumn{2}{l}{Level of C enrichment} &\colhead{$N$\tablenotemark{a}} }
\startdata
$\mbox{[C/Fe]}>0.5$ & dwarfs &    3   \\
                    & giants &   18   \\
$\mbox{[C/Fe]}>1.0$ & dwarfs &    2   \\
                    & giants &   11     
\enddata
\tablenotetext{a}{The values given for the dwarfs should be regarded as lower limits. See
  \S~\ref{carbon_ana} for further details.}
\end{deluxetable}

\pagestyle{empty}
\setlength{\voffset}{25mm}
\begin{deluxetable}{lrrrcccrlllccccrc}
\tabletypesize{\tiny} 
\tablecaption{\label{res_mp} The 286 bright metal-poor stars with $\mbox{[Fe/H]}<-1.0$}  
\tablewidth{0pt} 
\tablehead{ 
\colhead{Star name}&  
\colhead{$R.A.$}   &  
\colhead{$\delta$} &  
\colhead{$B$}      &  
\colhead{$(B-V)_{\rm \texttt{HP2}}$}    &  
\colhead{$(B-V)_{\rm J-K}$}    &  
\colhead{$E_{\rm (B-V)}$} & 
\colhead{$v_{\rm rad}$} & 
\colhead{\texttt{KP}} &  
\colhead{\texttt{HP2}}&  
\colhead{\texttt{GP}} &  
\colhead{$\mbox{[Fe/H]}_{\rm B-V}$} & 
\colhead{$\mbox{[Fe/H]}_{\texttt{\scriptsize HP2}}$}&  
\colhead{$\mbox{[Fe/H]}_{\rm Rossi}$}&  
\colhead{$\mbox{[Fe/H]}_{\rm final}$} & 
\colhead{$\mbox{[C/Fe]}$} &
\colhead{Redisc.}  \\
\colhead{(1)}&\colhead{(2)}&\colhead{(3)}&\colhead{(4)}&\colhead{(5)}&
\colhead{(6)}&\colhead{(7)}&\colhead{(8)}&\colhead{(9)}&\colhead{(10)}&
\colhead{(11)}&\colhead{(12)}&\colhead{(13)}&\colhead{(14)}&
\colhead{(15)}&\colhead{(16)}&\colhead{(17)} }
\startdata 
HE~0002$-$3233 & 00 05 32.3 & $-$32 16 36& 12.5& 0.42& 0.47& 0.02& 48& 1.30& 4.24& 0.48& $-$2.98 & $-$2.87& $-$2.67& $-$2.86& \nodata & *\\ 
HE~0003$-$0503 & 00 05 56.8 & $-$04 47 06& 11.4& 0.47& 0.82& 0.03& $-$2& 6.82& 3.39& 4.21& $-$2.28 & $-$1.03& $-$2.23& $-$2.23& 0.11 & \\ 
HE~0003$-$5106 & 00 06 29.8 & $-$50 49 31& 11.5& 0.69& 0.83& 0.01& 235& 6.82& 1.10& 1.73& $-$2.35 & $-$2.16& $-$2.26& $-$2.23& $-$0.67 & \\ 
HE~0007$-$1752 & 00 10 17.6 & $-$17 35 38& 12.2& 0.60& 0.65& 0.03& 216& 5.44& 1.63& 1.96& $-$2.24 & $-$2.12& $-$2.17& $-$2.16& $-$0.22 & *\\ 
HE~0010$-$1316 & 00 13 30.7 & $-$12 59 60& 13.0& 0.82& 1.02& 0.03& $-$86& 7.81& 0.72& 2.12& \nodata & $-$1.94& $-$2.44& $-$2.19& $-$0.69 & \\ 
HE~0012$-$5643 & 00 15 17.1 & $-$56 26 27& 12.1& 0.38& 0.43& 0.01& $-$285& 1.15& 4.51& 0.30& $-$3.15 & $-$3.00& $-$2.45& $-$2.94& \nodata & \\ 
HE~0013$-$0257 & 00 16 04.2 & $-$02 41 06& 13.7& 0.70& 0.75& 0.04& 36& 2.96& 1.24& 1.52& $-$3.37 & $-$3.07& $-$3.32& $-$3.21& 0.46 & \\ 
HE~0013$-$0522 & 00 16 28.1 & $-$05 05 52& 13.7& 0.64& 0.65& 0.03& $-$183& 3.56& 1.25& 2.02& $-$2.65 & $-$2.65& $-$2.83& $-$2.69& 0.43 & \\ 
HE~0015$-$0048 & 00 18 01.4 & 01 05 08& 14.0& 0.64& 0.78& 0.03& $-$46& 4.84& 1.25& 3.94& $-$2.93 & $-$2.38& $-$2.76& $-$2.61& 0.55 & \\ 
HE~0017$-$3646 & 00 20 26.1 & $-$36 30 20& 13.6& 0.54& 0.61& 0.01& 375& 3.88& 2.47& 0.62& $-$2.50 & $-$2.30& $-$2.58& $-$2.42& \nodata & *\\ 
HE~0017$-$5616 & 00 20 11.2 & $-$55 59 37& 10.5& 0.83& 0.64& 0.01& 30& 8.76& 0.55& 6.03& $-$0.88 & $-$1.44& $-$1.12& $-$1.12& 0.05 & \\ 
HE~0025$-$0223 & 00 28 28.3 & $-$02 06 42& 12.0& 0.38& 0.42& 0.03& 3& 4.90& 5.21& 1.49& $-$1.48 & $-$1.21& $-$1.57& $-$1.31& $-$0.31 & \\ 
HE~0027$-$1221 & 00 30 31.0 & $-$12 05 11& 13.7& 0.59& 0.50& 0.03& 14& 2.85& 1.71& 6.56& $-$2.40 & $-$2.56& $-$2.56& $-$2.56& 1.78 & *\\ 
HE~0030$-$5441 & 00 33 20.0 & $-$54 24 43& 11.1& 0.37& 0.42& 0.02& 0& 4.52& 5.03& 1.30& $-$1.58 & $-$1.01& $-$1.70& $-$1.20& $-$0.30 & \\ 
HE~0032$-$4056 & 00 34 33.7 & $-$40 39 30& 13.3& 0.42& 0.51& 0.02& $-$66& 1.63& 4.24& 0.39& $-$2.88 & $-$2.69& $-$2.84& $-$2.74& \nodata & \\ 
\enddata
\tablecomments{The table is available in its entirety only in electronic format. A portion
 is shown for guidance regarding its form and content.}
\end{deluxetable}

\setlength{\voffset}{0mm}
\begin{deluxetable}{lrrrcccrlllc}
\tabletypesize{\scriptsize} 
\tablecaption{\label{res_other} ``Non-metal-poor''objects}  
\tablewidth{0pt} 
\tablehead{ 
\colhead{Star name}&  
\colhead{$R.A.$}   &  
\colhead{$\delta$} &  
\colhead{$B$}      &  
\colhead{$(B-V)_{\rm \texttt{HP2}}$}    &  
\colhead{$(B-V)_{\rm J-K}$}    &  
\colhead{$E_{\rm (B-V)}$} & 
\colhead{$v_{\rm rad}$} & 
\colhead{\texttt{KP}} &  
\colhead{\texttt{HP2}}&  
\colhead{\texttt{GP}} &  
\colhead{Comm.} }  
\startdata 
HE~0000$-$3017 & 00 03 21.9 & $-$30 01 09& 10.0& 0.39& 0.42& 0.01& 19& 6.57& 4.99& 1.89& \\ 
HE~0000$-$4401 & 00 03 11.7 & $-$43 44 57& 11.3& 0.47& 0.45& 0.01& 23& 8.29& 3.60& 3.55& \\ 
HE~0000$-$5703 & 00 02 43.0 & $-$56 46 52& 8.4& \nodata& 0.27& 0.01& 40& 2.59& 9.37& 1.16& \\ 
HE~0001$-$4157 & 00 03 50.0 & $-$41 40 18& 11.5& 0.64& 0.53& 0.01& 21& 9.15& 1.73& 5.47& \\ 
HE~0001$-$4449 & 00 04 15.9 & $-$44 32 31& 11.7& 0.46& 0.45& 0.01& 27& 7.53& 3.53& 2.92& \\ 
HE~0001$-$5640 & 00 04 09.2 & $-$56 24 06& 10.4& 0.43& 0.43& 0.01& $-$15& 7.80& 4.06& 2.82& \\ 
HE~0002$-$3822 & 00 04 43.2 & $-$38 06 01& 11.3& 0.48& 0.49& 0.02& 42& 7.80& 3.39& 3.20& \\ 
HE~0002$-$5625 & 00 04 45.6 & $-$56 08 24& 13.0& 0.64& 0.59& 0.01& 60& 9.25& 1.65& 5.55& \\ 
HE~0003$-$2658 & 00 06 28.0 & $-$26 41 23& 11.5& 0.39& 0.38& 0.02& 2& 6.92& 4.95& 2.22& \\ 
HE~0004$-$1224 & 00 07 18.2 & $-$12 07 52& 10.4& 0.40& 0.48& 0.03& $-$22& 7.33& 4.59& 2.64& \\ 
HE~0004$-$1650 & 00 07 08.2 & $-$16 34 17& 11.4& 0.67& 0.53& 0.03& 75& 9.47& 1.44& 5.50& \\ 
HE~0005$-$4959 & 00 08 13.4 & $-$49 43 18& 10.7& 0.43& 0.43& 0.01& 22& 7.76& 4.01& 2.78& \\ 
HE~0005$-$5945 & 00 07 48.3 & $-$59 28 60& 10.9& 0.37& 0.35& 0.01& 29& 4.80& 6.50& 1.30& \\ 
HE~0006$-$0211 & 00 09 11.0 & $-$01 54 56& 11.8& 0.50& 0.46& 0.04& 4& 8.34& 3.02& 3.85& \\ 
HE~0006$-$1244 & 00 09 01.0 & $-$12 28 14& 11.0& 0.77& 0.60& 0.03& 45& 9.36& 0.84& 5.89& \\ 
\enddata
\tablecomments{HB=Horizontal Branch, He=He lines in the spectrum, Em=Emission lines in the spectrum, pec= , dist=distorted Ca K line, comp=possibly composited spectrum of two objects}
\tablecomments{The table is available in its entirety only in electronic format. A portion
 is shown for guidance regarding its form and content.}
\end{deluxetable}

\begin{deluxetable}{lcccccc}
\tablecaption{\label{feh_errors}Typical uncertainties in the [Fe/H] and [C/Fe] measurements} 
\tablewidth{0pt}
\tablehead{
\colhead{}& 
\colhead{$\sigma_{\rm {B-V}} (0.06)$}&\colhead{$\sigma_{\rm {J-K}}$ (0.03)}& \colhead{$\sigma_{\rm {KP}}$ (0.18)}& 
\colhead{$\sigma_{\rm {HP2}}$ (0.20)}& \colhead{$\sigma_{\rm {GP}}$ (0.15)}\\
\colhead{}& 
\colhead{(a)/(b)}&\colhead{(a)/(b)}& \colhead{(a)/(b)}& \colhead{(a)/(b)}& \colhead{(a)/(b)}}
\startdata 
$\mbox{[Fe/H]}_{\rm {B-V}}$  &0.17/0.45\tablenotemark{c}& \nodata &0.06/0.09& \nodata & \nodata \\
$\mbox{[Fe/H]}_{\rm {HP2}}$  & \nodata & \nodata &0.07/0.08&0.05/0.11\tablenotemark{d}& \nodata \\
$\mbox{[Fe/H]}_{\rm {Rossi}}$& \nodata &0.13/0.13&0.05/0.05& \nodata & \nodata \\
$\mbox{[C/Fe]}$              & \nodata & \nodata &0.06/0.04& \nodata &0.06/0.05
\enddata
\tablenotetext{a}{Derived from stars with $\mbox{[Fe/H]}<-2$.}
\tablenotetext{b}{Derived from stars with $\mbox{[Fe/H]}>-2$.}
\tablenotetext{c}{Values derived for dwarfs (i.e., $\texttt{HP2}>2.5$). For
  giants, the uncertainties are 0.25/0.50 based on $\sigma_{\rm {B-V}}=0.10$.}
\tablenotetext{d}{Values derived for dwarfs. For giants, the uncertainties are 0.15/0.13.}
\end{deluxetable}

\clearpage

\end{document}